\documentclass[reprint,superscriptaddress, amsmath,amssymb, aps]{revtex4-2}

\usepackage[utf8]{inputenc}
\usepackage[T1]{fontenc}
\usepackage{lineno}
\usepackage{graphicx}
\usepackage[dvipsnames]{xcolor}
\usepackage{todonotes}
\usepackage[normalem]{ulem}
\usepackage{xspace}
\usepackage{layouts}
\usepackage{siunitx}
\usepackage[hypertexnames=false]{hyperref}

\begin{document}

\renewcommand{\figurename}{Fig.}

\title{Si/SiGe QuBus for single electron information-processing devices with memory and micron-scale connectivity function}
\author{Ran Xue}
\author{Max Beer}
\author{Inga Seidler}
\affiliation{JARA-FIT Institute for Quantum Information, Forschungszentrum J\"ulich GmbH and RWTH Aachen University, Aachen, Germany}
\author{Simon Humpohl}
\affiliation{JARA-FIT Institute for Quantum Information, Forschungszentrum J\"ulich GmbH and RWTH Aachen University, Aachen, Germany}
\affiliation{ARQUE Systems GmbH, 52074 Aachen, Germany}
\author{Jhih-Sian Tu}
\author{Stefan Trellenkamp}
\affiliation{Helmholtz Nano Facility (HNF), Forschungszentrum J\"ulich, J\"ulich, Germany}
\author{Tom Struck}
\author{Hendrik Bluhm}
\author{Lars R. Schreiber}
\email{lars.schreiber@physik.rwth-aachen.de}
\affiliation{JARA-FIT Institute for Quantum Information, Forschungszentrum J\"ulich GmbH and RWTH Aachen University, Aachen, Germany}
\affiliation{ARQUE Systems GmbH, 52074 Aachen, Germany}

\begin{abstract}  
The connectivity within single carrier information-processing devices requires transport and storage of single charge quanta. Our all-electrical Si/SiGe shuttle device, called quantum bus (QuBus), spans a length of 10\,$\mathrm{\mu}$m and is operated by only six simply-tunable voltage pulses. It operates in conveyor-mode, i.e. the electron is adiabatically transported while confined to a moving QD. We introduce a characterization method, called shuttle-tomography, to benchmark the potential imperfections and local shuttle-fidelity of the QuBus. The fidelity of the single-electron shuttle across the full device and back (a total distance of 19\,$\mathrm{\mu}$m) is $(99.7 \pm 0.3)\,\%$. Using the QuBus, we position and detect up to 34 electrons and initialize a register of 34 quantum dots with arbitrarily chosen patterns of zero and single-electrons. The simple operation signals, compatibility with industry fabrication and low spin-environment-interaction in $^{28}$Si/SiGe, promises spin-conserving transport of spin qubits for quantum connectivity in quantum computing architectures.

\end{abstract}


\flushbottom
\maketitle

\thispagestyle{empty}
Controlling local charge densities in a semiconductor by metallic gate-electrodes sets the foundation of modern nanoelectronics. Raising their density triggered quantum mechanical effects paving the way to various nanoelectronic devices operating with single charge quanta. Discrete charge states of quantum dots (QDs) are stored to process digital information \cite{Chen1996} and the spin of individual electrons is used to encode quantum bits for quantum computing in semiconductors \cite{Ono2005, Hanson2007}. The exchange of charge quanta between functional blocks such as charge-photon interfaces \cite{Joecker2019,Holmes2020, Higginbottom2022}, quantum registers \cite{Volk2019}, spin manipulation zones \cite{Neumann2015}, single charge detectors \cite{Cassidy2007} and current standard devices \cite{Stein2017} would lead to quantum devices with new functionalities.

For conventional electronics, wires transport currents or voltages over extended distances. In quantum technology, wires cannot transport individual charges, as disorder limits their localization length hardly exceeding 100\,nm. In micron-sized quantum structures, the charging energy becomes impractically small for utilizing charge states with definite electron number. Here we present a device named quantum bus (QuBus) \cite{Seidler2022, Langrock2023}, which can solve this fundamental difficulty and might provide the key for the required \cite{Fowler2012, Gidney2021} scale-up of quantum computing architectures \cite{Vandersypen2017, Li2018, Boter2022, Kuenne2023}.

Single electron \cite{McNeil2011, Takada2019} and spin-conserving electron \cite{Jadot2021} shuttling has previously been demonstrated employing surface acoustic waves in piezoelectric semiconductors. Shuttling in non-polar materials such as silicon, which is highly attractive for quantum computing with electron-spins \cite{Xue2022, Noiri2022_Nat, Philips2022, Mills2022}, becomes more involved, as the electron transport requires a series of top gates. This additional complexity comes with the benefit of electron acceleration and velocity control \cite{Bertrand2016, Mills2019, Noiri2022_NatComm, Zwerver2022, Feng2023}. In particular, the conveyor-mode shuttling approach in Si/SiGe combines this advantage with predictable spin coherence during shuttling and the requirement for just four input signals independent from the length of the shuttle device \cite{Langrock2023}. High-fidelity short-range conveyor-mode charge and spin shuttling has been demonstrated \cite{Seidler2022, Struck2023}.

In this work, we all-electrically position and detect up to 34 electrons in a single-electron conveyor-mode QuBus in Si/SiGe. Despite its length of 10\,$\mathrm{\mu}$m and more than 100 electrostatic gates, the QuBus can be controlled by only six input terminals with low voltage pulse complexity. We introduce a characterization method we call shuttle tomography to benchmark the local shuttle fidelity of the QuBus using a single electron as a probe. By composing elementary pulses, we can control and detect any single electron pattern filling a series of 34 QDs. The conveyor-mode shuttle approach opens up new possibilities for probing local potential disorder in a quantum well, detecting single electrons with high lateral resolution across a length of 10\,$\mathrm{\mu}$m and boosting multi-electron control for scalable spin qubit quantum computation.

\section*{QuBus device and pulse segments} 

\begin{figure*}[htbp]
\centering
\includegraphics[width=0.95\textwidth]{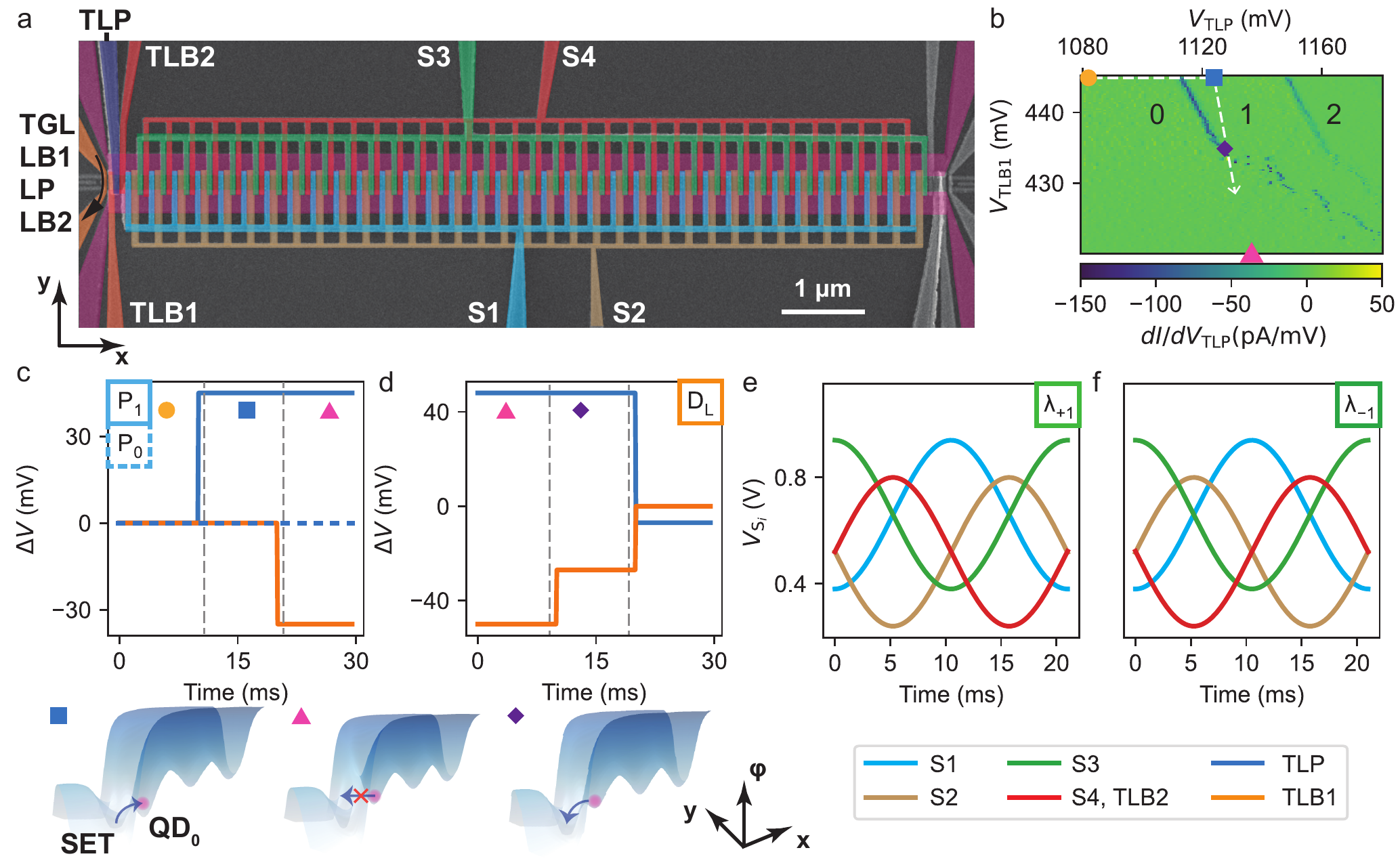}
\caption{\textbf{QuBus device and pulse segments.} \textbf{a}, False-colored scanning electron micrograph of a top-view on a device nominally identical to the measured device. The gate labelled TGL overlaps with two barrier gates LB1 and LB2 and accumulates the electron reservoirs for the left SET. The gate labeled TLB2 is electrically connected to terminal S$_4$. \textbf{b}, Charge stability diagram for controlling the $\mathrm{QD}_0$ filling (numbers) by individually pulsed terminals TLP and TLB1. Symbols indicate positions in gates space consistent with panels c and d. The electrostatic potential of the SET and $\mathrm{QD}_0$ are sketched for three positions beneath panels c, d. \textbf{c}, Voltage pulse segments $\Delta V$ for loading one (solid line) and zero electrons (dashed line) labelled P$_1$ and P$_0$, respectively. Colors refer to the input terminals TLP and TLB1 (see legend). \textbf{d}, Voltage pulse segment D$_L$ for detecting an electron in $\mathrm{QD}_0$ by the left SET. After detection the electron is unloaded to the SET. \textbf{e, f} Pulse segments $\mathrm{\lambda}_{+1}$ and $\mathrm{\lambda}_{-1}$ for shuttling an electron by a distance $\lambda$ in positive (panel e) and negative (panel f) x-direction using all terminals S$_i$ (see legend). All unchanged voltages during pulse segments are not plotted in panels c to f.} 
\label{fig:SEM image and pulses}
\end{figure*}

Our QuBus device consists of an undoped SiGe/Si/SiGe quantum well on top of which three electrically isolated metallic gate layers are fabricated by electron-beam lithography and metal lift-off (see the methods section for details on the device fabrication). The 10\,$\mathrm{\mu}$m long grounded split-gate on the first layer defines a nominally depleted one-dimensional electron channel (1DEC) in the quantum well. More than 100 clavier gates, equally distributed among the second and third layer above the 1DEC, enable the approximately uniform movement of single electrons (conveyor-mode shuttling) along the x-direction (Fig.\,\ref{fig:SEM image and pulses}a). Notably, every forth clavier gate is electrically connected to one of four gate sets S$_i$ ($i=1...4$).

\begin{figure*}[htbp]
\centering
\includegraphics[width=0.95\textwidth]{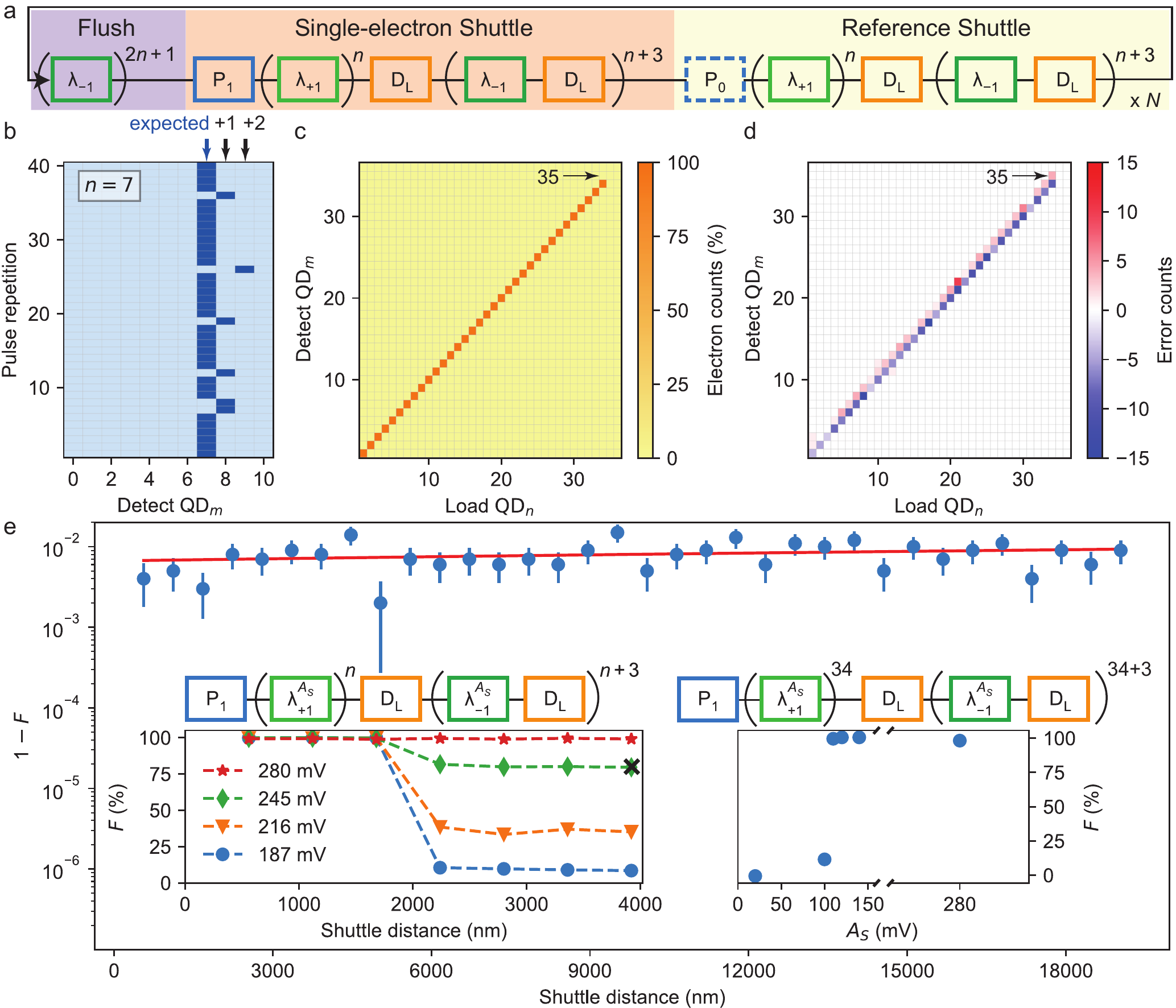}
\caption{\textbf{Single-charge shuttle tomography.} \textbf{a}, The pulse sequence of a shuttling tomography experiment consists of a flush pulse, a single-electron shuttle sequence and a reference shuttle pulse without an electron being loaded. The pulse is repeated $N$-times in order to gather statistics. \textbf{b}, Digitized electron-detection map recorded during the single-electron shuttle pulse for $A_S = 245$\,mV and $n=7$. By single-shot charge-detection (D$_L$) of the QD$_0$ to QD$_{10}$, we either find zero (light blue) or one (dark blue) electron for each of the 40 pulse cycles. The expected occurrence of an electron is indicated by the blue arrow and faulty locations by black arrows. The $1-F$ for this specific data set is marked by a cross in the left inset of panel e. \textbf{c}, Relative error counts i.e. fraction of electrons detected in QD$_m$, if one electron is loaded in QD$_n$ for $A_S = 280$\,mV and $N=1000$. For each column, we count all single-electron detection event from an electron-detection map as shown in panel b. \textbf{d}, Same as in panel c, but the error counts are plotted, i.e. the difference between expected and detected electron counts. The statistics shown  in panel c and d are based on the same data set. False electron detection of QD$_{35}$ is marked by an arrow. \textbf{e}, The shuttle infidelity $1-F$ (blue dots with 1$\sigma$ error bar) determined from the full shuttle-tomography sequence for $A_S = 280$\,mV and $N=1000$. Number of applied $\lambda_{\pm1}$-pulse segments is converted into total single-electron shuttle distance forwards and backwards. Red-line is fit to the data (see text). Left insert: Shuttle fidelity $F$ as a function of the total shuttle distance (forth and back) for various $A_S$ and $N=1000$ (lines are guide-to-the-eyes only). Right insert: Shuttle fidelity $F$ as a function of $A_S$ for maximum shuttle distance $n=34$. Note that $A_S$=280\,mV is only used for the forth $\lambda_{+1}$ and for the thirtieth $\lambda_{-1}$ shuttle pulses. The modified single electron shuttle pulses are sketched above the corresponding inserts. The applied flush and reference pulses are not shown for simplicity.}
\label{fig:single charge tomography}
\end{figure*}

On demand, a single electron can be loaded into the 1DEC from the left single-electron transistor (SET) formed by the gates TGL, LB1, LP and LB2. The plunger gate TLP of the leftmost quantum dot QD$_0$ controls the loading of exactly one electron from the SET to QD$_0$. The corresponding voltage pulses in gate space are indicated in Fig.\,\ref{fig:SEM image and pulses}b from the yellow dot to the blue square. This is followed by raising the tunnel barrier by gate TLB1 (pink triangle in Fig.\,\ref{fig:SEM image and pulses}b,c). We label the corresponding pulse segment as P$_1$. If the voltage $V_{\mathrm{TLP}}$ applied to gate TLP remains low during the entire segment, no electron is loaded which we label as P$_0$ (Fig.\,\ref{fig:SEM image and pulses}c). Reversely, we can also use the SET current $I$ to detect either zero or one  electron in QD$_0$ by the pulse segment $\mathrm{D}_\mathrm{L}$ (Fig.\,\ref{fig:SEM image and pulses}d). If an electron is detected, it is unloaded during the detection pulse (see Extended Data Fig.\,\ref{fig:read_hist} for details on the charge detection).

To shuttle the single electron in a moving QD, simple sinusoidal voltage pulses $V_{\mathrm{S}_i}(t)$ are applied to the gate sets S$_i$: 
\begin{equation}
V_{\mathrm{S}_i}(t) = A_S \cos\left(2\pi f t-\frac{\pi(i-1)}{2}\right)+B_s+\Delta B_s \frac{1+(-1)^i}{2},
\label{eq:voltages}
\end{equation}

\noindent where the pulse amplitude $A_S$ sets the confinement strength of the propagating sinusoidal potential created in the 1DEC. $B_S$ and $\Delta B_S$ are constant offsets for accumulating charges in the conduction band in the 1DEC, accommodating different distances of the gate sets from the 1DEC. The shuttle velocity is given by $f \lambda=14$\,$\mu$m$\cdot$s$^{-1}$, where the frequency of the shuttle pulse is $f=50$\,Hz and $\mathrm{\lambda}= 280$\,nm is the lateral period of the potential in the 1DEC. 

To transport all electrons in the 1DEC by a distance of $\lambda$ in the positive (negative) x-direction, the pulse segment $\mathrm{\lambda}_{+1}$ ($\mathrm{\lambda}_{-1}$) is employed (Fig.\,\ref{fig:SEM image and pulses}e,f)). Note that all voltages applied to gates of the device return to their initial values at the end of each $\mathrm{\lambda}_{\pm1}$ pulse segment. This implies that the correction of the SET’s operating point for capacitive cross-coupling to the clavier gate sets S$_i$ is constant and thus simple. The right SET and the rightmost clavier gates are not used here and voltages are chosen to have an open 1DEC towards an energetically lower lying electron reservoir. Hence, in total only six voltage pulses $V_{\mathrm{S}_i} $, $V_{\mathrm{TLP}}$ and $V_{\mathrm{TLB1}}$ given by the elementary pulse segments P$_0$, P$_1$, $\mathrm{\lambda}_{+1}$, $\mathrm{\lambda}_{-1}$ and $\mathrm{D}_\mathrm{L}$ control the whole 10\,$\mathrm{\mu}$m long shuttle device, inside of which a total of 35 QDs (QD$_i$ with $i=0...34$) are formed along the 1DEC.

\section*{Single-charge shuttle tomography} 

In order to discuss the composition and interplay of pulse segments during the operation of the QuBus, we choose the pulse sequence called shuttle tomography as a first example. The sequence is designed to measure the local shuttle fidelity $F_{\mathrm{\lambda}}$, i.e. the shuttle success rate of $\mathrm{\lambda}_{+1}$ for a specific position of the probe electron. Thus, we might identify local weak spots in the QuBus, although charge detection by the SET is limited to QD$_0$. The strategy is to shuttle a single electron from the left end of 1DEC further into the 1DEC by some short distance and then back to the detector. We repeat this experiment in order to record the shuttle fidelity and sequentially increase the shuttle distance until the electron is shuttled the full distance of 19\,$\mathrm{\mu}$m forth and back. The obtained data serves as a benchmark of the local shuttle fidelity in the QuBus. The corresponding pulse sequence is displayed in Fig.\,\ref{fig:single charge tomography}a. First, the depleted 1DEC is loaded with a single electron (P$_1$) which is then shuttled into the 1DEC for a distance of $n \cdot \lambda$ (by repeating $\mathrm{\lambda}_{+1}$ $n$-times). Afterwards the electron filling of the first $n+4$ QDs is measured by consecutively detecting ($\mathrm{D}_\mathrm{L}$) and shuttling one period back towards the SET ($\mathrm{\lambda}_{-1}$). Finally, we apply a reference pulse by repeating the full pulse, but replace the P$_1$ segment by P$_0$. This shuttle pulse is $N$-times repeated.

As an instructive subset of such a measurement over $N=40$ pulse repetitions with $n=7$ and $A_S=245$\,mV, we observe the filling of each of the first eleven QDs as shown in Fig.\,\ref{fig:single charge tomography}b. During the majority of shuttle pulses the electron remains within QD$_7$, into which it was loaded. This result indicates a well operating QuBus. Sometimes the electron is detected in QD$_8$ and QD$_9$, thus the shuttle process failed during these repetitions. Via the reference pulse segment, we check whether electrons leak into the 1DEC. Since we never observe any electrons during the reference pulse across thousands of repetitions for all $n$, we conclude that there is no such leakage and the SET charge-detector does not faultily detect electrons in an empty QD. 

The full observation of shuttle tomography with $A_S=280$\,mV, $N=1000$ and $n=1...36$ shows that the single electron is nearly always detected in the expected QD$_n$, into which it has been loaded (Fig.\,\ref{fig:single charge tomography}c and Extended Data Fig.\,\ref{fig:Shuttle_tomo_1e}). Additionally, no electrons are observed for $n=35, 36$. This is expected as the right end of the 1DEC is open and the 1DEC only contains QD$_0$ to QD$_{34}$. Hence, the electron is pushed out of the 1DEC through its right end for $n=35, 36$.

We introduce the electron count $C_m^l$  to express the number of electrons detected in QD$_{m}$ summing over all $N$ pulse repetitions where $l$ is the expected filling of QD\,$_{m}$, which is 1 only for $n=m$ and 0 otherwise. Thus, the error count of each QD relative to its expectation is given by $\Delta C_m^l = C_m^l - l\cdot N$. The single-electron error count (Fig.\,\ref{fig:single charge tomography}d) reveals that in very few repetitions the electron was detected in a QD$_m$ with $m>n$ and almost never for $m<n$. In approximately 1\,\% of the repetitions the electrons seem to disappear ($\sum_{m,l} C_m^l<N$) (see Extended Data Fig.\,\ref{fig:Shuttle_tomo_1e}). Remarkably, some electrons are detected in QD$_{35}$ when loaded in QD$_{34}$, although QD$_{35}$ does not exist. Hence, delayed electrons got stuck during a $\mathrm{\lambda}_{-1}$ pulse segment, instead of hopping over one QD during a $\mathrm{\lambda}_{1}$ pulse segment. This indicates a directionality of the shuttle error.

We define one shuttle pulse as successful, if three conditions are simultaneously fulfilled: (I) An electron is detected in the $n$-th QD, into which an electron has been loaded. (II) No electron is detected in all other QDs, which are detected during the sequence. (III) No electron is detected during the reference shuttle sequence in any QD. We count the number of successful shuttle pulses with the same $n$ and divide by the total number $N$ of pulse repetitions to get the charge shuttle-fidelity $F(n)$.

With $A_S=280$\,mV and $N=1000$ for each of the $n=1...34$ covering a shuttle distance of $2 n \cdot280$\,nm, we observe an average shuttle infidelity of $1-F= (0.785 \pm 0.051)$\,\% (Fig.\,\ref{fig:single charge tomography}e). This infidelity, however, also includes errors from initialisation and detection pulse segments. Remarkably, $F(n)$ is almost independent of the shuttle distance. Therefore, we split the observed infidelity into two error sources: First, the shuttle error $\varepsilon_{\mathrm{\lambda}}$ occurring during each $\mathrm{\lambda}_{\pm1}$-pulse, which is a shuttle dependent error that accumulates over the increment of shuttle periods. Second, an electron loading and detection (LD) error $\varepsilon_{\mathrm{LD}}$, which is independent of shuttling and attributed to errors occurring during the P$_1$ (no electron initialized by error) and $\mathrm{D}_\mathrm{L}$ (no electron detected by error) pulse segments. We linearly fit $\ln(F) = A\cdot n + B$ where $A = 2 \ln (F_{\mathrm{\lambda}}) = 2 \ln (1-\varepsilon_{\mathrm{\lambda}}), B = \ln (F_{\mathrm{LD}}) = \ln (1-\varepsilon_{\mathrm{LD}})$ and find the average shuttle fidelity per period $F_{\mathrm{\lambda}} = (99.996 \pm 0.003)\,\%$ at $A_S=280$\,mV corresponding to an expected orbital splitting of 4\,meV in $\mathrm{QD}_1$ (see Extended Data Fig.\,\ref{fig:Sim_linecuts}). The LD error is  $\varepsilon_{\mathrm{LD}}= (0.7 \pm 0.1)\,\%$, thus the LD-corrected shuttle fidelity across the full channel and back is $\hat{F}(34) = (99.7 \pm 0.3)\,\%$ (total distance 19\,$\mathrm{\mu}$m, see methods section for details on the estimations of errors).

Finally, we provoke shuttling errors by reducing $A_S$ and thus the confinement of the QDs in the shuttle potential. Note that the amplitude of the flush pulse is always constant at $A_S=280$\,mV. We observe that as we decrease $A_S$, the shuttle fidelity drops between the third and the forth shuttle period and then remains constant (left insert in Fig.\,\ref{fig:single charge tomography}e). Thus, we attribute the decrease in $F$ to a local weak spot in the QuBus potential, likely due to static potential disorder. To confirm this hypothesis, we modify the shuttling tomography pulse sequence by tuning $A_S$ as a function of shuttle distance. Therefore, temporarily enhanced confinement is realized by keeping $A_S = 280$\,mV during the fourth $\mathrm{\lambda}_{+1}$ and the $(n-4)$-th $\mathrm{\lambda}_{-1}$ pulse segment, thus at the position of the weak spot only. This demonstrates a tunable method to shuttle electrons over the QuBus with high $F$ at much lower $A_S$ applied during all other pulse segments $\mathrm{\lambda}_{\pm1}$ (right inset of Fig.\,\ref{fig:single charge tomography}e). The observed cut-off amplitude at $100$\,mV matches well with simulations of semiconductor-oxide interface charge-defect induced potential disorder in the 1DEC \cite{Langrock2023}. The origin of the weak spot in the QuBus requires further investigation. Note that for the measurement of $F(n)$, we cannot fully exclude two errors appearing during shuttling which compensate each other. However, the observation that faulty shuttling behavior occurs locally in the QuBus, makes it probable that two such spots should be separately observed by the $n$-dependence of the shuttling tomography.

\section*{Multi-electron operation}

\begin{figure}[ht]
\centering
\includegraphics[width=\linewidth]{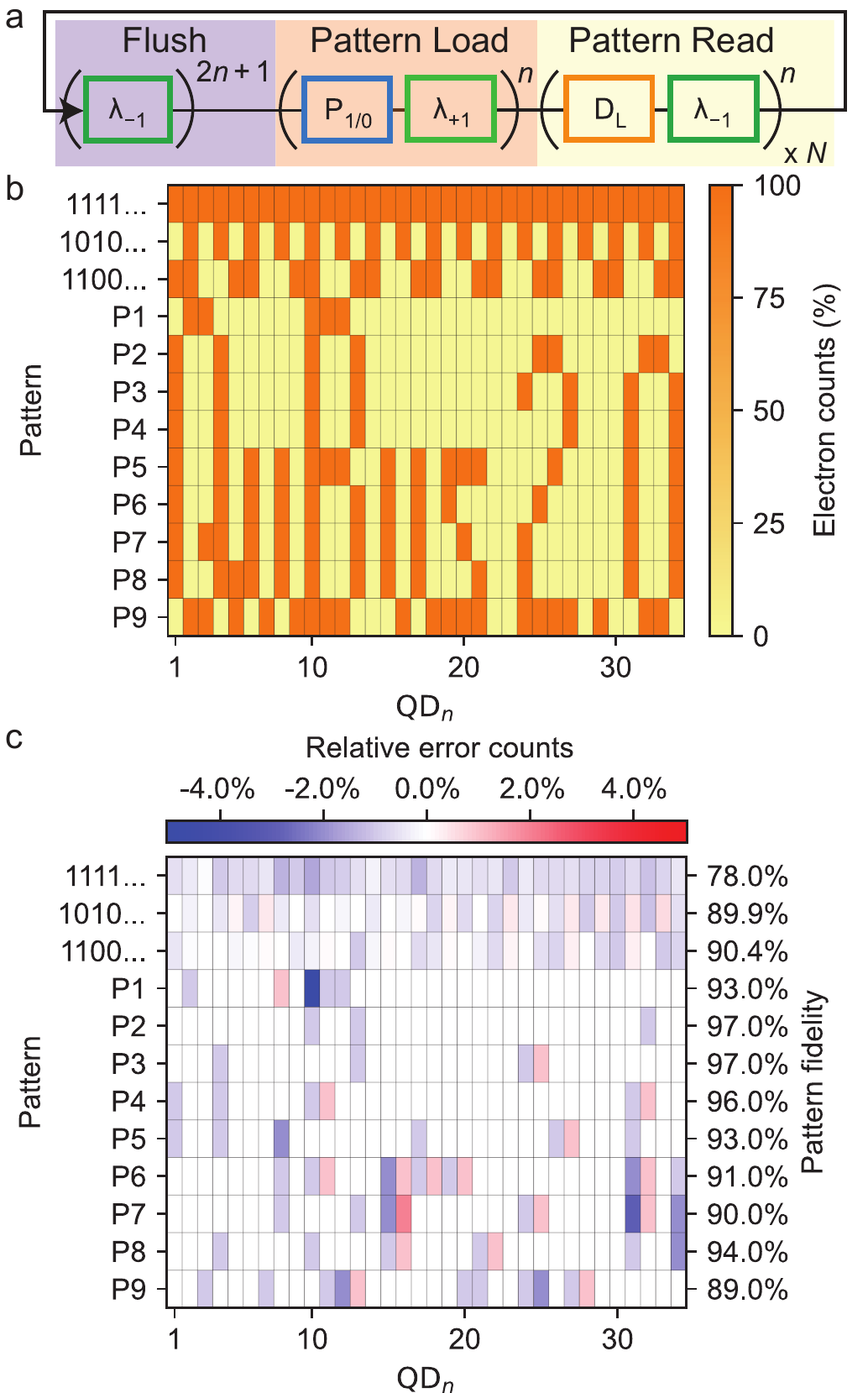}
\caption{\textbf{Multi-electron operation.} \textbf{a}, Pulse sequence composed of elementary pulse elements. \textbf{b}, Detected number of electrons in QD$_n$ normalized by the total number of pulse repetitions $N$ for an arbitrary target filling pattern (rows with labels on the left). For P$_i$ the target pattern is non-periodic but can be identified from the data due to the low shuttle error and is therefore not explicitly given. \textbf{c}, The relative error counts of each QD for shuttled patterns. The pattern fidelity, as defined in the text, is shown on the right for each individual row.}
\label{fig:pattern shuttling}
\end{figure}

For the shuttle tomography, only exactly one electron was loaded into the QuBus at a time. The QuBus can also be operated with many electrons using the aforementioned elementary pulse segments. Each electron can be placed in any of the QDs between QD$_1$ and QD$_{34}$ in a controlled manner. Thus, we can create a pattern of electron fillings in a 34 QD register. The pulse sequence for loading and detecting an arbitrary electron pattern in the QuBus (Fig.\,\ref{fig:pattern shuttling}a) is similar to the sequence employed during shuttle tomography. The repetition of $\mathrm{\lambda}_{+1}$ segments is replaced by a series of single $\mathrm{\lambda}_{+1}$ interleaved with  P$_0$ and P$_1$ pulse segments. The latter determine the pattern filling the QDs. The key expectation is that any $\mathrm{\lambda}_{\pm1}$ segment should move all electrons simultaneously by shifting the sinusoidal potential in the 1DEC.

Using $A_S=280$\,mV, we load one electron in each of the 34 QDs ($1111...$), every second QD ($1010...$) or a more complex periodic pattern ($1100...$). We repeat the pattern loading and detecting for $N=1000$ times to gather statistics on the electron count in each QD (Fig.\,\ref{fig:pattern shuttling}b). We observe that the fraction of counted electrons in all QDs is very close to the expected filling pattern. Next, nine non-periodic patterns P1-P9, representing the lines of a binary image comprising $34\times9$ bits, are successfully loaded and detected as observed from the statistics of $N=100$ pulse repetitions. 

The dominant bluish color in the error-count map for all patterns (Fig.\,\ref{fig:pattern shuttling}c) reveals that the main error is the apparent loss of  electrons. We assign this notion to the dominance of the initialisation and detection error $\varepsilon_{\mathrm{LD}}$. It also explains why the pattern fidelity, which we define analogue to the shuttle-fidelity as the rate of successfully and exclusively placing and detecting electrons in all intended QDs, is lowest for the pattern with the highest electron count (111...). Blue/red dipoles in the error-count map indicate a shuttle error. As for shuttle tomography, we mainly observe individual electrons being misplaced by one QD to the right, provided this adjacent QD is nominally empty. This observation underlines the directional character of the shuttle error, which we already noted for the shuttle tomography.

\section*{Conclusion}   

Our all-electrical Si/SiGe electron-shuttle device is successfully operated in conveyor mode. Only six input terminals control the more than 100 clavier gates of the $10$\,µm long device. Independent of its length, only four sinusoidal signals are required to operate the shuttle as well as two signals for loading and detecting electrons by a single-electron transistor. We introduce a method called shuttle tomography, which uses a single electron to probe the local shuttle fidelity and thus local imperfections in the confinement of the moving QD. We estimate the fidelity for shuttling one electron across the full length and back, thus a total distance of 19\,$\mu$m, to be $\hat{F} = (99.7 \pm 0.3)\,\%$. Employing other pulse sequences composed of the five elementary pulses for our QuBus, we programmatically distribute and detect up to 34 electrons across the 34 QDs formed in the shuttle device. Any QD filling pattern can be initialized and we encode a digital image, the pixels of which are represented by single electrons.

The QDs can be interpreted as a 34 bit stack with a maximum of 34 electrons or as the initialization procedure of a series of 34 QDs towards a quantum register for spin qubits. Preparing such patterns of electrons in a one-dimensional channel also opens up possibilities to study the interplay of tunnel coupling and Coulomb interaction for a specific charge configuration. Furthermore, our conveyor-mode QuBus device paves the way to scalable quantum computation, since it is expected that the electron-spin evolution is deterministic during conveyor-mode shuttling at a velocity of approximately 8\,m$\cdot$\,s$^{-1}$ and spin-coherent shuttle fidelities of 99.9 \% are predicted \cite{Langrock2023}. Notably our QuBus is technologically compatible to industrial fabrication and Si/SiGe has been proven to be an ideal host-crystal for spin qubits. Spin-qubit connectivity across a distance of several micrometers could be a game changer for spin quantum computation. 

\section*{Methods}

\subsection*{The QuBus device}
The undoped quantum well heterostructure is grown by chemical vapour deposition on a 200\,mm silicon wafer and consists of a 7\,nm tensile-strained silicon layer sandwiched between two relaxed layers of $\text{Si}_{0.70}\text{Ge}_{0.30}$. The upper barrier layer of $\text{Si}_{0.70}\text{Ge}_{0.30}$ has a nominal thickness of 30\,nm and is capped by 2\,nm of Si. Ohmic contacts to the quantum well are created by the selective phosphorus ion-implantation followed by a rapid thermal anneal at \SI{700}{\celsius} for \SI{30}{\second}. The contacts are then metalized using optical lithography and metal lift-off. Three metallic gate layers including fan-out are fabricated via electron beam lithography and evaporation followed by metal lift-off. A scanning electron micrograph of a device nominally identical to the device measured in this work can be seen in Fig. \ref{fig:SEM image and pulses}a. The first gate layer is deposited directly onto the silicon capping layer, the native oxide layer of which was removed immediately before metal evaporation via HF etching. For this lowest layer \SI{15}{\nano\meter} of palladium is used in order to fabricate a suitable metal-semiconductor junction. The later two gate layers are fabricated on \SI{7}{\nano\meter} of atomic layer deposited $\text{Al}_2\text{O}_3$ and consist of \SI{5}{\nano\meter} of titanium and \si{22}/\SI{29}{\nano\meter} of platinum for the second and third layer, respectively.

The first fine gate layer defines both the SET plunger and barrier gates as well as the channel-confining split-gate. The split-gate constrains the 1DEC to a width below \SI{200}{\nano\meter}. The second and third metal gate layers define the SET top gates as well as the clavier gates, which form the individual QDs in the 1DEC. These clavier gates have a width of \SI{60}{\nano\meter} with a pitch of \SI{70}{\nano\meter}. The designed distance between SET and QD$_0$ and thus the tunnel-coupling is based on Ref. \cite{Klos2018}.   

\subsection*{Experimental setup}
Experiments are conducted in an Oxford Triton 200 dilution refrigerator at approximately $60$\,mK. Voltage pulses are generated by a Zurich Instruments HDAWG8 and superposed with DC voltages from a home-built DAC by a passive voltage adder at room-temperature. All signal lines are filtered by pi-filters with a cut-off frequency of 1\,kHz. No low-temperature filtering is used. The SET current is converted by the low-noise transimpedance amplifier SP983c from Basel Precision Instruments with a cut-off frequency of 3\,kHz and digitized by an AlazarTech ATS9440 waveform digitizer. The composition of pulse sequences employs the open source python package qupulse \cite{Humpohl2023}.

\subsection*{Error Estimation}

Here we discuss and estimate the error probability for manipulating the charge state during each elementary pulse segment $\varepsilon_i$ with $i=\mathrm{P}_0, \mathrm{P}_1, \mathrm{\lambda}_{+1}, \mathrm{\lambda}_{-1}, \mathrm{D}_\mathrm{L}$. First, we assume for simplicity that the average error for shuttling one electron by a distance of $\lambda$ is $\varepsilon_\mathrm{\lambda} \approx \varepsilon_{\mathrm{\lambda}_{+1}} \approx \varepsilon_{\mathrm{\lambda}_{-1}}$ despite the experimentally observed small directionality. Since we never observe any electrons during the reference pulse across thousands of shuttle tomography repetitions for all $n$, we conclude that $\varepsilon_{\mathrm{P}_0} \approx 0$ and that the detector does not faultily detect electrons in an empty QD. We combine the error from loading one electron $\varepsilon_{\mathrm{P}_1}$ and missing an electron during detection $\varepsilon_{\mathrm{D}_\mathrm{L}}$ to be the loading and detection error $\varepsilon_{\mathrm{LD}}$ with $(1-\varepsilon_{\mathrm{LD}})=(1-\varepsilon_{\mathrm{P}_1}) (1-\varepsilon_{\mathrm{D}_\mathrm{L}})$. The experimentally observed shuttle fidelity $F(n)$ during a shuttle tomography pulse sequence of shuttle distance $2n\lambda$ is composed of several elementary pulse segments: 

\begin{equation*}
F(n) = (1-\varepsilon_{\mathrm{LD}})\cdot (1-\varepsilon_{\mathrm{\lambda}})^{2n}.
\end{equation*}

By linearly fitting $\ln(F(n))$, we find $\varepsilon_{\mathrm{LD}}= (0.7 \pm 0.1)\,\%$ and the average shuttle fidelity per period $F_{\mathrm{\lambda}} = 1-\varepsilon_\mathrm{\lambda}= (99.996 \pm 0.003)\,\%$. Thus, the expected LD-corrected shuttle fidelity $\hat{F}(n) = (1-\varepsilon_{\mathrm{\lambda}})^{2n}$ for a total shuttle distance of $\approx 19$\,$\mathrm{\mu}$m is $\hat{F}(34)=(1-\varepsilon_\mathrm{\lambda})^{68} = (99.7 \pm 0.3)\,\%$. This corresponds to the fidelity of shuttling a single electron across the full QuBus and back.

\section*{Acknowledgements}
 We thank Łukasz Cywiński for helpful comments on the manuscript. This work has been funded by the German Research Foundation (DFG) under Germany's Excellence Strategy - Cluster of Excellence Matter and Light for Quantum Computing" (ML4Q) EXC 2004/1 - 390534769 and by the Federal Ministry of Education and Research under Contract No. FKZ: 13N14778. Project Si-QuBus received funding from the QuantERA ERA-NET Cofund in Quantum Technologies implemented within the European Union's Horizon 2020 Program. The device fabrication has been done at HNF - Helmholtz Nano Facility, Research Center Juelich GmbH \cite{albrecht_hnf_2017}. 

\section*{Author contributions}
R.X. conducted the experiments with M.B and L.R.S. R.X. and M.B. analysed the data with I.S. and L.R.S. S.H. provided measurement software. T.S prepared the passive voltage adder. M.B. adapted the developed process by R.X, J.-S.T and I.S for the device fabrication and fabricated the device. S.T. operated e-beam lithography. L.R.S. designed the device and supervised the experiment. L.R.S and H.B. provided guidance to all authors. R.X., M.B. and L.R.S. wrote the manuscript, which was commented by all other authors.

\section*{Competing interests}
R.X., I.S., H.B. and L.R.S. are co-inventors of patent applications that cover conveyor-mode shuttling and its applications. L.R.S. and H.B. are founders and shareholders of ARQUE Systems GmbH. The other authors declare no competing interests.

\newpage
\appendix
\setcounter{figure}{0}
\renewcommand{\figurename}{Extended Data Fig.}

\begin{figure*}[ht]
    \centering
    \includegraphics[width=0.95\textwidth]{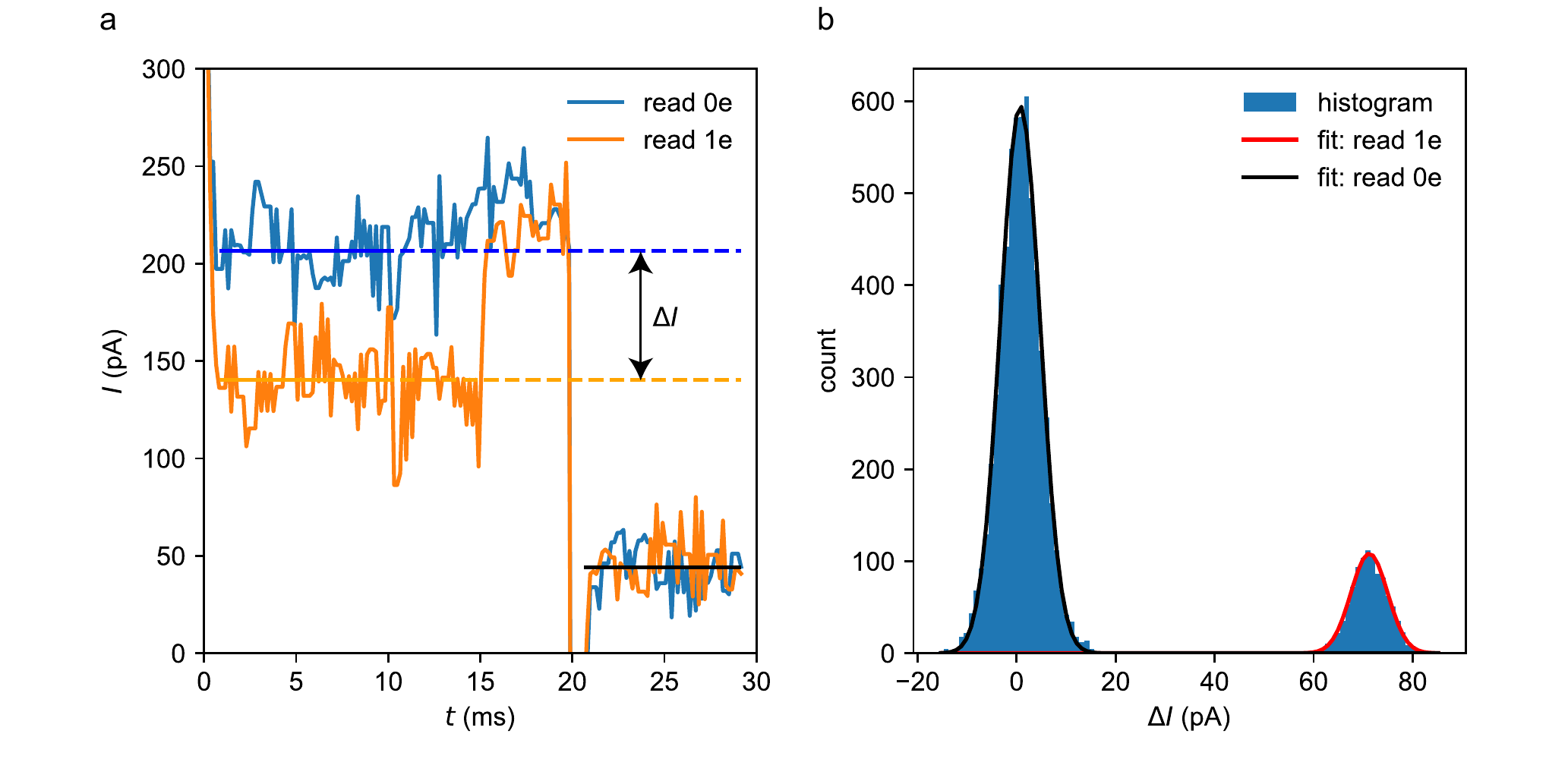}
    \caption{\textbf{Single-shot charge-readout.} \textbf{a}, SET current trace $I$ during the detection pulse segment D$\mathrm{_L}$ in which both readouts of electron present and no electron in QD$_0$ are illustrated in orange and blue respectively. The solid horizontal lines indicate the averaged current levels of corresponding read-out segment from which $\Delta I$ is calculated. \textbf{b}, Histogram of evaluated current differences $\Delta I$ (as defined in panel a) for  1000 detection pulses. Binning results by two gaussian functions indicates the statistics of detecting an empty QD$_0$ (black) or an electron within QD$_0$ (red), accordingly. Due to the repetition of reference shuttling pulses without presence of an electron, the events of detecting an empty QD$_0$ dominates compared to detecting a single electron as intended. The overlap between the two gaussian fits is relevant to the detection error and is evaluated to $2.29\time10^{-20}$ which indicates a negligible estimation error.}
    \label{fig:read_hist}
\end{figure*}

\newpage
\begin{figure*}[ht]
    \centering
    \includegraphics[width=0.95\textwidth]{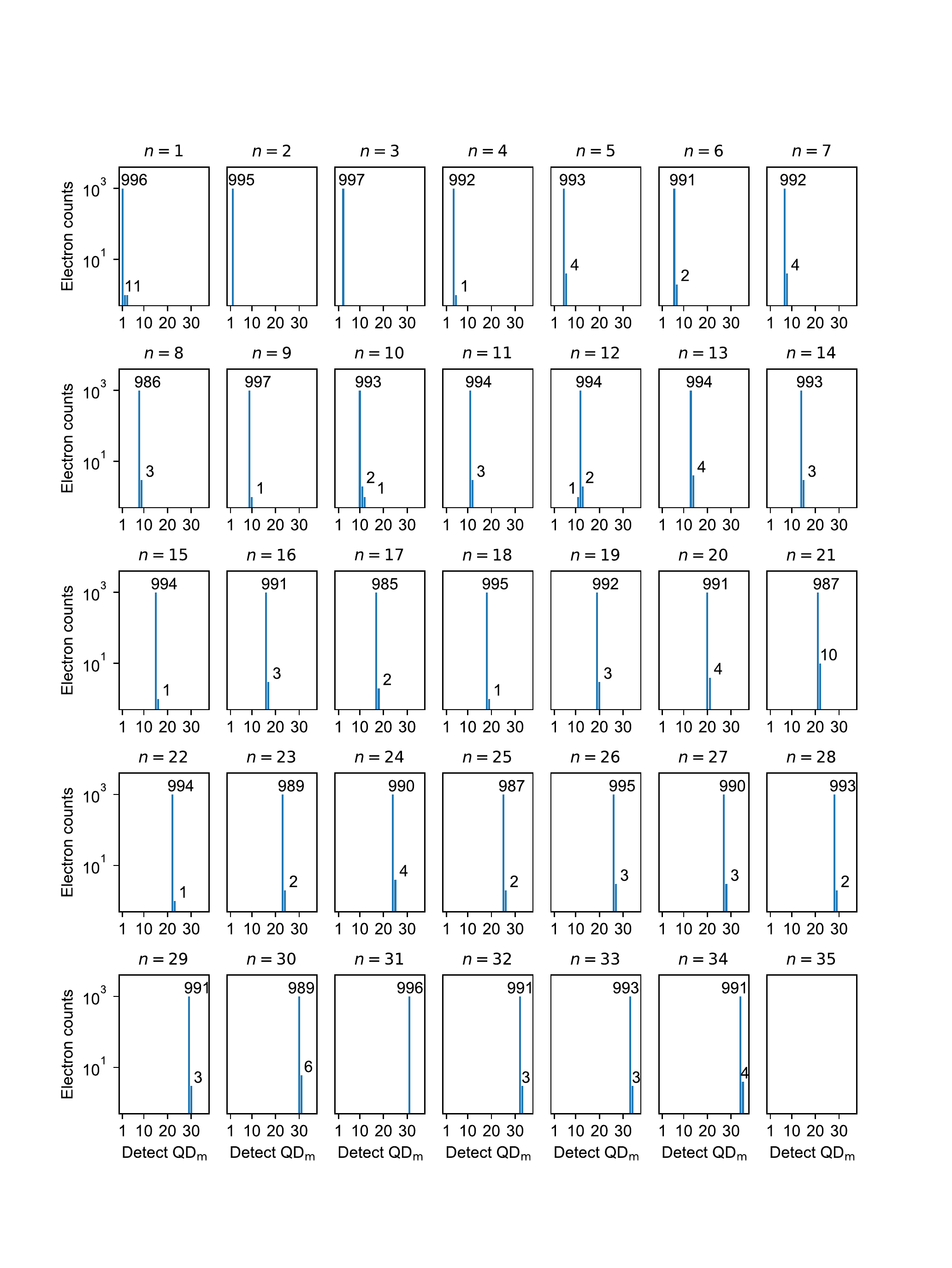}
    \caption{\textbf{Electron counts for single charge shuttle-tomography.} Each panel is the electron count for $N=1000$ shuttle-pulse repetitions for $A_\mathrm{S}=280$\,meV as a function of the QD number $m$, in which the electrons are detected. For each panel a different $n$ is used in the pulse sequence. Numbers label the height of bars larger than zero. For $n=35$ the shuttle distance exceeds the physical dimension of QuBus.}
    \label{fig:Shuttle_tomo_1e}
\end{figure*}

\newpage
\begin{figure*}[ht]
    \centering
    \includegraphics[width=0.95\textwidth]{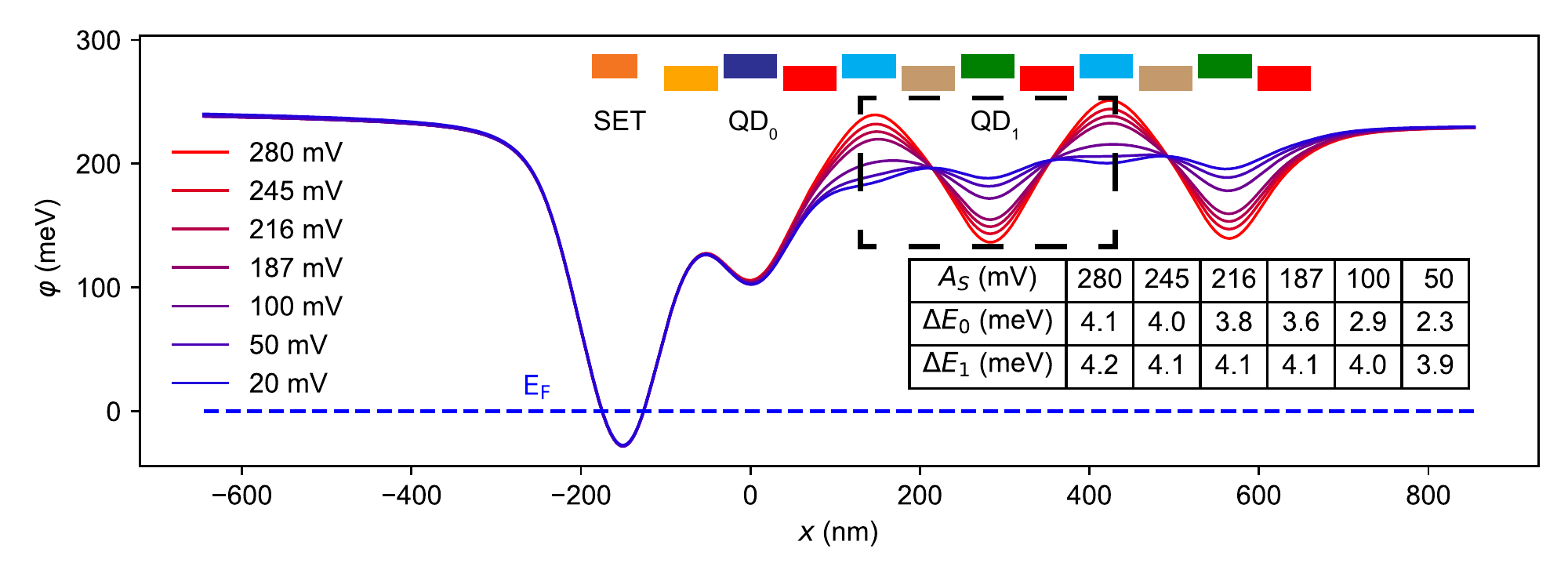}
    \caption{\textbf{Simulated electrostatic potential of 1DEC.} Line-cut of the finite-element numerically simulated electrostatic energy $\varphi(x)$ along the center of QuBus 1DEC for various $A_S$ in the area of the left SET and QD$_{0...2}$. Thomas-Fermi approximation is used to simulate the SET in order to take screening of electron reservoirs into account. The Schr\"odinger-Poisson equation is solved for QD$_1$ in the region marked by the black dashed rectangle. Three lowest orbitals \text{s}, \text{$p_x$} and \text{$p_y$} are used to calculate the lowest orbital splittings $\Delta E_0$ and $\Delta E_1$. The dashed horizontal line represents the Fermi energy $E_\mathrm{F}$ for electrons confined by the Si/SiGe heterostructure at operating temperature. Cross sections of SET top-gate and clavier gates are shown on top of the simulated potential using the same color scheme as in Fig.\,\ref{fig:SEM image and pulses}.}
    \label{fig:Sim_linecuts}
\end{figure*}


\begin{thebibliography}{34}%
\makeatletter
\providecommand \@ifxundefined [1]{%
 \@ifx{#1\undefined}
}%
\providecommand \@ifnum [1]{%
 \ifnum #1\expandafter \@firstoftwo
 \else \expandafter \@secondoftwo
 \fi
}%
\providecommand \@ifx [1]{%
 \ifx #1\expandafter \@firstoftwo
 \else \expandafter \@secondoftwo
 \fi
}%
\providecommand \natexlab [1]{#1}%
\providecommand \enquote  [1]{``#1''}%
\providecommand \bibnamefont  [1]{#1}%
\providecommand \bibfnamefont [1]{#1}%
\providecommand \citenamefont [1]{#1}%
\providecommand \href@noop [0]{\@secondoftwo}%
\providecommand \href [0]{\begingroup \@sanitize@url \@href}%
\providecommand \@href[1]{\@@startlink{#1}\@@href}%
\providecommand \@@href[1]{\endgroup#1\@@endlink}%
\providecommand \@sanitize@url [0]{\catcode `\\12\catcode `\$12\catcode
  `\&12\catcode `\#12\catcode `\^12\catcode `\_12\catcode `\%12\relax}%
\providecommand \@@startlink[1]{}%
\providecommand \@@endlink[0]{}%
\providecommand \url  [0]{\begingroup\@sanitize@url \@url }%
\providecommand \@url [1]{\endgroup\@href {#1}{\urlprefix }}%
\providecommand \urlprefix  [0]{URL }%
\providecommand \Eprint [0]{\href }%
\providecommand \doibase [0]{https://doi.org/}%
\providecommand \selectlanguage [0]{\@gobble}%
\providecommand \bibinfo  [0]{\@secondoftwo}%
\providecommand \bibfield  [0]{\@secondoftwo}%
\providecommand \translation [1]{[#1]}%
\providecommand \BibitemOpen [0]{}%
\providecommand \bibitemStop [0]{}%
\providecommand \bibitemNoStop [0]{.\EOS\space}%
\providecommand \EOS [0]{\spacefactor3000\relax}%
\providecommand \BibitemShut  [1]{\csname bibitem#1\endcsname}%
\let\auto@bib@innerbib\@empty
\bibitem [{\citenamefont {Chen}\ \emph {et~al.}(1996)\citenamefont {Chen},
  \citenamefont {Korotkov},\ and\ \citenamefont {Likharev}}]{Chen1996}%
  \BibitemOpen
  \bibfield  {author} {\bibinfo {author} {\bibfnamefont {R.~H.}\ \bibnamefont
  {Chen}}, \bibinfo {author} {\bibfnamefont {A.~N.}\ \bibnamefont {Korotkov}},\
  and\ \bibinfo {author} {\bibfnamefont {K.~K.}\ \bibnamefont {Likharev}},\
  }\bibfield  {title} {\bibinfo {title} {{Single‐electron transistor
  logic}},\ }\href {https://doi.org/10.1063/1.115637} {\bibfield  {journal}
  {\bibinfo  {journal} {Appl. Phys. Lett.}\ }\textbf {\bibinfo {volume} {68}},\
  \bibinfo {pages} {1954} (\bibinfo {year} {1996})}\BibitemShut {NoStop}%
\bibitem [{\citenamefont {Ono}\ \emph {et~al.}(2005)\citenamefont {Ono},
  \citenamefont {Fujiwara}, \citenamefont {Nishiguchi}, \citenamefont
  {Inokawa},\ and\ \citenamefont {Takahashi}}]{Ono2005}%
  \BibitemOpen
  \bibfield  {author} {\bibinfo {author} {\bibfnamefont {Y.}~\bibnamefont
  {Ono}}, \bibinfo {author} {\bibfnamefont {A.}~\bibnamefont {Fujiwara}},
  \bibinfo {author} {\bibfnamefont {K.}~\bibnamefont {Nishiguchi}}, \bibinfo
  {author} {\bibfnamefont {H.}~\bibnamefont {Inokawa}},\ and\ \bibinfo {author}
  {\bibfnamefont {Y.}~\bibnamefont {Takahashi}},\ }\bibfield  {title} {\bibinfo
  {title} {Manipulation and detection of single electrons for future
  information processing},\ }\href {https://doi.org/10.1063/1.1843271}
  {\bibfield  {journal} {\bibinfo  {journal} {J. Appl. Phys.}\ }\textbf
  {\bibinfo {volume} {97}},\ \bibinfo {pages} {031101} (\bibinfo {year}
  {2005})}\BibitemShut {NoStop}%
\bibitem [{\citenamefont {Hanson}\ \emph {et~al.}(2007)\citenamefont {Hanson},
  \citenamefont {Kouwenhoven}, \citenamefont {Petta}, \citenamefont {Tarucha},\
  and\ \citenamefont {Vandersypen}}]{Hanson2007}%
  \BibitemOpen
  \bibfield  {author} {\bibinfo {author} {\bibfnamefont {R.}~\bibnamefont
  {Hanson}}, \bibinfo {author} {\bibfnamefont {L.~P.}\ \bibnamefont
  {Kouwenhoven}}, \bibinfo {author} {\bibfnamefont {J.~R.}\ \bibnamefont
  {Petta}}, \bibinfo {author} {\bibfnamefont {S.}~\bibnamefont {Tarucha}},\
  and\ \bibinfo {author} {\bibfnamefont {L.~M.~K.}\ \bibnamefont
  {Vandersypen}},\ }\bibfield  {title} {\bibinfo {title} {Spins in few-electron
  quantum dots},\ }\href {https://doi.org/10.1103/RevModPhys.79.1217}
  {\bibfield  {journal} {\bibinfo  {journal} {Rev. Mod. Phys.}\ }\textbf
  {\bibinfo {volume} {79}},\ \bibinfo {pages} {1217} (\bibinfo {year}
  {2007})}\BibitemShut {NoStop}%
\bibitem [{\citenamefont {Joecker}\ \emph {et~al.}(2019)\citenamefont
  {Joecker}, \citenamefont {Cerfontaine}, \citenamefont {Haupt}, \citenamefont
  {Schreiber}, \citenamefont {Kardyna\l{}},\ and\ \citenamefont
  {Bluhm}}]{Joecker2019}%
  \BibitemOpen
  \bibfield  {author} {\bibinfo {author} {\bibfnamefont {B.}~\bibnamefont
  {Joecker}}, \bibinfo {author} {\bibfnamefont {P.}~\bibnamefont
  {Cerfontaine}}, \bibinfo {author} {\bibfnamefont {F.}~\bibnamefont {Haupt}},
  \bibinfo {author} {\bibfnamefont {L.~R.}\ \bibnamefont {Schreiber}}, \bibinfo
  {author} {\bibfnamefont {B.~E.}\ \bibnamefont {Kardyna\l{}}},\ and\ \bibinfo
  {author} {\bibfnamefont {H.}~\bibnamefont {Bluhm}},\ }\bibfield  {title}
  {\bibinfo {title} {Transfer of a quantum state from a photonic qubit to a
  gate-defined quantum dot},\ }\href
  {https://doi.org/10.1103/PhysRevB.99.205415} {\bibfield  {journal} {\bibinfo
  {journal} {Phys. Rev. B}\ }\textbf {\bibinfo {volume} {99}},\ \bibinfo
  {pages} {205415} (\bibinfo {year} {2019})}\BibitemShut {NoStop}%
\bibitem [{\citenamefont {Arakawa}\ and\ \citenamefont
  {Holmes}(2020)}]{Holmes2020}%
  \BibitemOpen
  \bibfield  {author} {\bibinfo {author} {\bibfnamefont {Y.}~\bibnamefont
  {Arakawa}}\ and\ \bibinfo {author} {\bibfnamefont {M.~J.}\ \bibnamefont
  {Holmes}},\ }\bibfield  {title} {\bibinfo {title} {{Progress in quantum-dot
  single photon sources for quantum information technologies: A broad spectrum
  overview}},\ }\href {https://doi.org/10.1063/5.0010193} {\bibfield  {journal}
  {\bibinfo  {journal} {Appl. Phys. Rev.}\ }\textbf {\bibinfo {volume} {7}},\
  \bibinfo {pages} {021309} (\bibinfo {year} {2020})}\BibitemShut {NoStop}%
\bibitem [{\citenamefont {Higginbottom}\ \emph {et~al.}(2022)\citenamefont
  {Higginbottom}, \citenamefont {Kurkjian}, \citenamefont {Chartrand},
  \citenamefont {Kazemi}, \citenamefont {Brunelle}, \citenamefont
  {{MacQuarrie}}, \citenamefont {Klein}, \citenamefont {Lee-Hone},
  \citenamefont {Stacho}, \citenamefont {Ruether}, \citenamefont {Bowness},
  \citenamefont {Bergeron}, \citenamefont {{DeAbreu}}, \citenamefont
  {Harrigan}, \citenamefont {Kanaganayagam}, \citenamefont {Marsden},
  \citenamefont {Richards}, \citenamefont {Stott}, \citenamefont {Roorda},
  \citenamefont {Morse}, \citenamefont {Thewalt},\ and\ \citenamefont
  {Simmons}}]{Higginbottom2022}%
  \BibitemOpen
  \bibfield  {author} {\bibinfo {author} {\bibfnamefont {D.~B.}\ \bibnamefont
  {Higginbottom}}, \bibinfo {author} {\bibfnamefont {A.~T.~K.}\ \bibnamefont
  {Kurkjian}}, \bibinfo {author} {\bibfnamefont {C.}~\bibnamefont {Chartrand}},
  \bibinfo {author} {\bibfnamefont {M.}~\bibnamefont {Kazemi}}, \bibinfo
  {author} {\bibfnamefont {N.~A.}\ \bibnamefont {Brunelle}}, \bibinfo {author}
  {\bibfnamefont {E.~R.}\ \bibnamefont {{MacQuarrie}}}, \bibinfo {author}
  {\bibfnamefont {J.~R.}\ \bibnamefont {Klein}}, \bibinfo {author}
  {\bibfnamefont {N.~R.}\ \bibnamefont {Lee-Hone}}, \bibinfo {author}
  {\bibfnamefont {J.}~\bibnamefont {Stacho}}, \bibinfo {author} {\bibfnamefont
  {M.}~\bibnamefont {Ruether}}, \bibinfo {author} {\bibfnamefont
  {C.}~\bibnamefont {Bowness}}, \bibinfo {author} {\bibfnamefont
  {L.}~\bibnamefont {Bergeron}}, \bibinfo {author} {\bibfnamefont
  {A.}~\bibnamefont {{DeAbreu}}}, \bibinfo {author} {\bibfnamefont {S.~R.}\
  \bibnamefont {Harrigan}}, \bibinfo {author} {\bibfnamefont {J.}~\bibnamefont
  {Kanaganayagam}}, \bibinfo {author} {\bibfnamefont {D.~W.}\ \bibnamefont
  {Marsden}}, \bibinfo {author} {\bibfnamefont {T.~S.}\ \bibnamefont
  {Richards}}, \bibinfo {author} {\bibfnamefont {L.~A.}\ \bibnamefont {Stott}},
  \bibinfo {author} {\bibfnamefont {S.}~\bibnamefont {Roorda}}, \bibinfo
  {author} {\bibfnamefont {K.~J.}\ \bibnamefont {Morse}}, \bibinfo {author}
  {\bibfnamefont {M.~L.~W.}\ \bibnamefont {Thewalt}},\ and\ \bibinfo {author}
  {\bibfnamefont {S.}~\bibnamefont {Simmons}},\ }\bibfield  {title} {\bibinfo
  {title} {Optical observation of single spins in silicon},\ }\href
  {https://doi.org/10.1038/s41586-022-04821-y} {\bibfield  {journal} {\bibinfo
  {journal} {Nature}\ }\textbf {\bibinfo {volume} {607}},\ \bibinfo {pages}
  {266} (\bibinfo {year} {2022})}\BibitemShut {NoStop}%
\bibitem [{\citenamefont {Volk}\ \emph {et~al.}(2019)\citenamefont {Volk},
  \citenamefont {Zwerver}, \citenamefont {Mukhopadhyay}, \citenamefont
  {Eendebak}, \citenamefont {van Diepen}, \citenamefont {Dehollain},
  \citenamefont {Hensgens}, \citenamefont {Fujita}, \citenamefont {Reichl},
  \citenamefont {Wegscheider},\ and\ \citenamefont {Vandersypen}}]{Volk2019}%
  \BibitemOpen
  \bibfield  {author} {\bibinfo {author} {\bibfnamefont {C.}~\bibnamefont
  {Volk}}, \bibinfo {author} {\bibfnamefont {A.~M.~J.}\ \bibnamefont
  {Zwerver}}, \bibinfo {author} {\bibfnamefont {U.}~\bibnamefont
  {Mukhopadhyay}}, \bibinfo {author} {\bibfnamefont {P.~T.}\ \bibnamefont
  {Eendebak}}, \bibinfo {author} {\bibfnamefont {C.~J.}\ \bibnamefont {van
  Diepen}}, \bibinfo {author} {\bibfnamefont {J.~P.}\ \bibnamefont
  {Dehollain}}, \bibinfo {author} {\bibfnamefont {T.}~\bibnamefont {Hensgens}},
  \bibinfo {author} {\bibfnamefont {T.}~\bibnamefont {Fujita}}, \bibinfo
  {author} {\bibfnamefont {C.}~\bibnamefont {Reichl}}, \bibinfo {author}
  {\bibfnamefont {W.}~\bibnamefont {Wegscheider}},\ and\ \bibinfo {author}
  {\bibfnamefont {L.~M.~K.}\ \bibnamefont {Vandersypen}},\ }\bibfield  {title}
  {\bibinfo {title} {Loading a quantum-dot based ``qubyte'' register},\ }\href
  {https://doi.org/10.1038/s41534-019-0146-y} {\bibfield  {journal} {\bibinfo
  {journal} {npj Quantum Inf}\ }\textbf {\bibinfo {volume} {5}},\ \bibinfo
  {pages} {29} (\bibinfo {year} {2019})}\BibitemShut {NoStop}%
\bibitem [{\citenamefont {Neumann}\ and\ \citenamefont
  {Schreiber}(2015)}]{Neumann2015}%
  \BibitemOpen
  \bibfield  {author} {\bibinfo {author} {\bibfnamefont {R.}~\bibnamefont
  {Neumann}}\ and\ \bibinfo {author} {\bibfnamefont {L.~R.}\ \bibnamefont
  {Schreiber}},\ }\bibfield  {title} {\bibinfo {title} {{Simulation of
  micro-magnet stray-field dynamics for spin qubit manipulation}},\ }\href
  {https://doi.org/10.1063/1.4921291} {\bibfield  {journal} {\bibinfo
  {journal} {J. Appl. Phys.}\ }\textbf {\bibinfo {volume} {117}},\ \bibinfo
  {pages} {193903} (\bibinfo {year} {2015})}\BibitemShut {NoStop}%
\bibitem [{\citenamefont {Cassidy}\ \emph {et~al.}(2007)\citenamefont
  {Cassidy}, \citenamefont {Dzurak}, \citenamefont {Clark}, \citenamefont
  {Petersson}, \citenamefont {Farrer}, \citenamefont {Ritchie},\ and\
  \citenamefont {Smith}}]{Cassidy2007}%
  \BibitemOpen
  \bibfield  {author} {\bibinfo {author} {\bibfnamefont {M.~C.}\ \bibnamefont
  {Cassidy}}, \bibinfo {author} {\bibfnamefont {A.~S.}\ \bibnamefont {Dzurak}},
  \bibinfo {author} {\bibfnamefont {R.~G.}\ \bibnamefont {Clark}}, \bibinfo
  {author} {\bibfnamefont {K.~D.}\ \bibnamefont {Petersson}}, \bibinfo {author}
  {\bibfnamefont {I.}~\bibnamefont {Farrer}}, \bibinfo {author} {\bibfnamefont
  {D.~A.}\ \bibnamefont {Ritchie}},\ and\ \bibinfo {author} {\bibfnamefont
  {C.~G.}\ \bibnamefont {Smith}},\ }\bibfield  {title} {\bibinfo {title}
  {Single shot charge detection using a radio-frequency quantum point
  contact},\ }\href {https://doi.org/10.1063/1.2809370} {\bibfield  {journal}
  {\bibinfo  {journal} {Appl. Phys. Lett.}\ }\textbf {\bibinfo {volume} {91}},\
  \bibinfo {pages} {222104} (\bibinfo {year} {2007})}\BibitemShut {NoStop}%
\bibitem [{\citenamefont {Stein}\ \emph {et~al.}(2017)\citenamefont {Stein},
  \citenamefont {Scherer}, \citenamefont {Gerster}, \citenamefont {Behr},
  \citenamefont {Götz}, \citenamefont {Pesel}, \citenamefont {Leicht},
  \citenamefont {Ubbelohde}, \citenamefont {Weimann}, \citenamefont {Pierz},
  \citenamefont {Schumacher},\ and\ \citenamefont {Hohls}}]{Stein2017}%
  \BibitemOpen
  \bibfield  {author} {\bibinfo {author} {\bibfnamefont {F.}~\bibnamefont
  {Stein}}, \bibinfo {author} {\bibfnamefont {H.}~\bibnamefont {Scherer}},
  \bibinfo {author} {\bibfnamefont {T.}~\bibnamefont {Gerster}}, \bibinfo
  {author} {\bibfnamefont {R.}~\bibnamefont {Behr}}, \bibinfo {author}
  {\bibfnamefont {M.}~\bibnamefont {Götz}}, \bibinfo {author} {\bibfnamefont
  {E.}~\bibnamefont {Pesel}}, \bibinfo {author} {\bibfnamefont
  {C.}~\bibnamefont {Leicht}}, \bibinfo {author} {\bibfnamefont
  {N.}~\bibnamefont {Ubbelohde}}, \bibinfo {author} {\bibfnamefont
  {T.}~\bibnamefont {Weimann}}, \bibinfo {author} {\bibfnamefont
  {K.}~\bibnamefont {Pierz}}, \bibinfo {author} {\bibfnamefont {H.~W.}\
  \bibnamefont {Schumacher}},\ and\ \bibinfo {author} {\bibfnamefont
  {F.}~\bibnamefont {Hohls}},\ }\bibfield  {title} {\bibinfo {title}
  {Robustness of single-electron pumps at sub-ppm current accuracy level},\
  }\href {https://doi.org/10.1088/1681-7575/54/1/S1} {\bibfield  {journal}
  {\bibinfo  {journal} {Metrologia}\ }\textbf {\bibinfo {volume} {54}},\
  \bibinfo {pages} {S1} (\bibinfo {year} {2017})}\BibitemShut {NoStop}%
\bibitem [{\citenamefont {Seidler}\ \emph {et~al.}(2022)\citenamefont
  {Seidler}, \citenamefont {Struck}, \citenamefont {Xue}, \citenamefont
  {Focke}, \citenamefont {Trellenkamp}, \citenamefont {Bluhm},\ and\
  \citenamefont {Schreiber}}]{Seidler2022}%
  \BibitemOpen
  \bibfield  {author} {\bibinfo {author} {\bibfnamefont {I.}~\bibnamefont
  {Seidler}}, \bibinfo {author} {\bibfnamefont {T.}~\bibnamefont {Struck}},
  \bibinfo {author} {\bibfnamefont {R.}~\bibnamefont {Xue}}, \bibinfo {author}
  {\bibfnamefont {N.}~\bibnamefont {Focke}}, \bibinfo {author} {\bibfnamefont
  {S.}~\bibnamefont {Trellenkamp}}, \bibinfo {author} {\bibfnamefont
  {H.}~\bibnamefont {Bluhm}},\ and\ \bibinfo {author} {\bibfnamefont {L.~R.}\
  \bibnamefont {Schreiber}},\ }\bibfield  {title} {\bibinfo {title}
  {Conveyor-mode single-electron shuttling in {Si}/{SiGe} for a scalable
  quantum computing architecture},\ }\href
  {https://doi.org/10.1038/s41534-022-00615-2} {\bibfield  {journal} {\bibinfo
  {journal} {npj Quantum Inf}\ }\textbf {\bibinfo {volume} {8}},\ \bibinfo
  {pages} {100} (\bibinfo {year} {2022})}\BibitemShut {NoStop}%
\bibitem [{\citenamefont {Langrock}\ \emph {et~al.}(2023)\citenamefont
  {Langrock}, \citenamefont {Krzywda}, \citenamefont {Focke}, \citenamefont
  {Seidler}, \citenamefont {Schreiber},\ and\ \citenamefont
  {Cywi\ifmmode~\acute{n}\else \'{n}\fi{}ski}}]{Langrock2023}%
  \BibitemOpen
  \bibfield  {author} {\bibinfo {author} {\bibfnamefont {V.}~\bibnamefont
  {Langrock}}, \bibinfo {author} {\bibfnamefont {J.~A.}\ \bibnamefont
  {Krzywda}}, \bibinfo {author} {\bibfnamefont {N.}~\bibnamefont {Focke}},
  \bibinfo {author} {\bibfnamefont {I.}~\bibnamefont {Seidler}}, \bibinfo
  {author} {\bibfnamefont {L.~R.}\ \bibnamefont {Schreiber}},\ and\ \bibinfo
  {author} {\bibfnamefont {L.}~\bibnamefont {Cywi\ifmmode~\acute{n}\else
  \'{n}\fi{}ski}},\ }\bibfield  {title} {\bibinfo {title} {Blueprint of a
  scalable spin qubit shuttle device for coherent mid-range qubit transfer in
  disordered {Si}/{SiGe}/{SiO}$_{2}$},\ }\href
  {https://doi.org/10.1103/PRXQuantum.4.020305} {\bibfield  {journal} {\bibinfo
   {journal} {PRX Quantum}\ }\textbf {\bibinfo {volume} {4}},\ \bibinfo {pages}
  {020305} (\bibinfo {year} {2023})}\BibitemShut {NoStop}%
\bibitem [{\citenamefont {Fowler}\ \emph {et~al.}(2012)\citenamefont {Fowler},
  \citenamefont {Mariantoni}, \citenamefont {Martinis},\ and\ \citenamefont
  {Cleland}}]{Fowler2012}%
  \BibitemOpen
  \bibfield  {author} {\bibinfo {author} {\bibfnamefont {A.~G.}\ \bibnamefont
  {Fowler}}, \bibinfo {author} {\bibfnamefont {M.}~\bibnamefont {Mariantoni}},
  \bibinfo {author} {\bibfnamefont {J.~M.}\ \bibnamefont {Martinis}},\ and\
  \bibinfo {author} {\bibfnamefont {A.~N.}\ \bibnamefont {Cleland}},\
  }\bibfield  {title} {\bibinfo {title} {Surface codes: Towards practical
  large-scale quantum computation},\ }\href
  {https://doi.org/10.1103/PhysRevA.86.032324} {\bibfield  {journal} {\bibinfo
  {journal} {Phys. Rev. A}\ }\textbf {\bibinfo {volume} {86}},\ \bibinfo
  {pages} {032324} (\bibinfo {year} {2012})}\BibitemShut {NoStop}%
\bibitem [{\citenamefont {Gidney}\ and\ \citenamefont
  {Eker{\aa{}}}(2021)}]{Gidney2021}%
  \BibitemOpen
  \bibfield  {author} {\bibinfo {author} {\bibfnamefont {C.}~\bibnamefont
  {Gidney}}\ and\ \bibinfo {author} {\bibfnamefont {M.}~\bibnamefont
  {Eker{\aa{}}}},\ }\bibfield  {title} {\bibinfo {title} {How to factor 2048
  bit {RSA} integers in 8 hours using 20 million noisy qubits},\ }\href
  {https://doi.org/10.22331/q-2021-04-15-433} {\bibfield  {journal} {\bibinfo
  {journal} {{Quantum}}\ }\textbf {\bibinfo {volume} {5}},\ \bibinfo {pages}
  {433} (\bibinfo {year} {2021})}\BibitemShut {NoStop}%
\bibitem [{\citenamefont {Vandersypen}\ \emph {et~al.}(2017)\citenamefont
  {Vandersypen}, \citenamefont {Bluhm}, \citenamefont {Clarke}, \citenamefont
  {Dzurak}, \citenamefont {Ishihara}, \citenamefont {Morello}, \citenamefont
  {Reilly}, \citenamefont {Schreiber},\ and\ \citenamefont
  {Veldhorst}}]{Vandersypen2017}%
  \BibitemOpen
  \bibfield  {author} {\bibinfo {author} {\bibfnamefont {L.~M.~K.}\
  \bibnamefont {Vandersypen}}, \bibinfo {author} {\bibfnamefont
  {H.}~\bibnamefont {Bluhm}}, \bibinfo {author} {\bibfnamefont {J.~S.}\
  \bibnamefont {Clarke}}, \bibinfo {author} {\bibfnamefont {A.~S.}\
  \bibnamefont {Dzurak}}, \bibinfo {author} {\bibfnamefont {R.}~\bibnamefont
  {Ishihara}}, \bibinfo {author} {\bibfnamefont {A.}~\bibnamefont {Morello}},
  \bibinfo {author} {\bibfnamefont {D.~J.}\ \bibnamefont {Reilly}}, \bibinfo
  {author} {\bibfnamefont {L.~R.}\ \bibnamefont {Schreiber}},\ and\ \bibinfo
  {author} {\bibfnamefont {M.}~\bibnamefont {Veldhorst}},\ }\bibfield  {title}
  {\bibinfo {title} {Interfacing spin qubits in quantum dots and donors—hot,
  dense, and coherent},\ }\href {https://doi.org/10.1038/s41534-017-0038-y}
  {\bibfield  {journal} {\bibinfo  {journal} {npj Quantum Inf}\ }\textbf
  {\bibinfo {volume} {3}},\ \bibinfo {pages} {34} (\bibinfo {year}
  {2017})}\BibitemShut {NoStop}%
\bibitem [{\citenamefont {Li}\ \emph {et~al.}(2018)\citenamefont {Li},
  \citenamefont {Petit}, \citenamefont {Franke}, \citenamefont {Dehollain},
  \citenamefont {Helsen}, \citenamefont {Steudtner}, \citenamefont {Thomas},
  \citenamefont {Yoscovits}, \citenamefont {Singh}, \citenamefont {Wehner},
  \citenamefont {Vandersypen}, \citenamefont {Clarke},\ and\ \citenamefont
  {Veldhorst}}]{Li2018}%
  \BibitemOpen
  \bibfield  {author} {\bibinfo {author} {\bibfnamefont {R.}~\bibnamefont
  {Li}}, \bibinfo {author} {\bibfnamefont {L.}~\bibnamefont {Petit}}, \bibinfo
  {author} {\bibfnamefont {D.~P.}\ \bibnamefont {Franke}}, \bibinfo {author}
  {\bibfnamefont {J.~P.}\ \bibnamefont {Dehollain}}, \bibinfo {author}
  {\bibfnamefont {J.}~\bibnamefont {Helsen}}, \bibinfo {author} {\bibfnamefont
  {M.}~\bibnamefont {Steudtner}}, \bibinfo {author} {\bibfnamefont {N.~K.}\
  \bibnamefont {Thomas}}, \bibinfo {author} {\bibfnamefont {Z.~R.}\
  \bibnamefont {Yoscovits}}, \bibinfo {author} {\bibfnamefont {K.~J.}\
  \bibnamefont {Singh}}, \bibinfo {author} {\bibfnamefont {S.}~\bibnamefont
  {Wehner}}, \bibinfo {author} {\bibfnamefont {L.~M.~K.}\ \bibnamefont
  {Vandersypen}}, \bibinfo {author} {\bibfnamefont {J.~S.}\ \bibnamefont
  {Clarke}},\ and\ \bibinfo {author} {\bibfnamefont {M.}~\bibnamefont
  {Veldhorst}},\ }\bibfield  {title} {\bibinfo {title} {A crossbar network for
  silicon quantum dot qubits},\ }\href
  {https://www.science.org/doi/abs/10.1126/sciadv.aar3960} {\bibfield
  {journal} {\bibinfo  {journal} {Sci. Adv.}\ }\textbf {\bibinfo {volume}
  {4}},\ \bibinfo {pages} {eaar3960} (\bibinfo {year} {2018})}\BibitemShut
  {NoStop}%
\bibitem [{\citenamefont {Boter}\ \emph {et~al.}(2022)\citenamefont {Boter},
  \citenamefont {Dehollain}, \citenamefont {van Dijk}, \citenamefont {Xu},
  \citenamefont {Hensgens}, \citenamefont {Versluis}, \citenamefont {Naus},
  \citenamefont {Clarke}, \citenamefont {Veldhorst}, \citenamefont
  {Sebastiano},\ and\ \citenamefont {Vandersypen}}]{Boter2022}%
  \BibitemOpen
  \bibfield  {author} {\bibinfo {author} {\bibfnamefont {J.~M.}\ \bibnamefont
  {Boter}}, \bibinfo {author} {\bibfnamefont {J.~P.}\ \bibnamefont
  {Dehollain}}, \bibinfo {author} {\bibfnamefont {J.~P.}\ \bibnamefont {van
  Dijk}}, \bibinfo {author} {\bibfnamefont {Y.}~\bibnamefont {Xu}}, \bibinfo
  {author} {\bibfnamefont {T.}~\bibnamefont {Hensgens}}, \bibinfo {author}
  {\bibfnamefont {R.}~\bibnamefont {Versluis}}, \bibinfo {author}
  {\bibfnamefont {H.~W.}\ \bibnamefont {Naus}}, \bibinfo {author}
  {\bibfnamefont {J.~S.}\ \bibnamefont {Clarke}}, \bibinfo {author}
  {\bibfnamefont {M.}~\bibnamefont {Veldhorst}}, \bibinfo {author}
  {\bibfnamefont {F.}~\bibnamefont {Sebastiano}},\ and\ \bibinfo {author}
  {\bibfnamefont {L.~M.}\ \bibnamefont {Vandersypen}},\ }\bibfield  {title}
  {\bibinfo {title} {Spiderweb array: A sparse spin-qubit array},\ }\href
  {https://doi.org/10.1103/PhysRevApplied.18.024053} {\bibfield  {journal}
  {\bibinfo  {journal} {Phys. Rev. Applied}\ }\textbf {\bibinfo {volume}
  {18}},\ \bibinfo {pages} {024053} (\bibinfo {year} {2022})}\BibitemShut
  {NoStop}%
\bibitem [{\citenamefont {K\"unne}\ \emph {et~al.}(2023)\citenamefont
  {K\"unne}, \citenamefont {Willmes}, \citenamefont {Oberl\"ander},
  \citenamefont {Gorjaew}, \citenamefont {Teske}, \citenamefont {Bhardwaj},
  \citenamefont {Beer}, \citenamefont {Kammerloher}, \citenamefont {Otten},
  \citenamefont {Seidler}, \citenamefont {Xue}, \citenamefont {Schreiber},\
  and\ \citenamefont {Bluhm}}]{Kuenne2023}%
  \BibitemOpen
  \bibfield  {author} {\bibinfo {author} {\bibfnamefont {M.}~\bibnamefont
  {K\"unne}}, \bibinfo {author} {\bibfnamefont {A.}~\bibnamefont {Willmes}},
  \bibinfo {author} {\bibfnamefont {M.}~\bibnamefont {Oberl\"ander}}, \bibinfo
  {author} {\bibfnamefont {C.}~\bibnamefont {Gorjaew}}, \bibinfo {author}
  {\bibfnamefont {J.~D.}\ \bibnamefont {Teske}}, \bibinfo {author}
  {\bibfnamefont {H.}~\bibnamefont {Bhardwaj}}, \bibinfo {author}
  {\bibfnamefont {M.}~\bibnamefont {Beer}}, \bibinfo {author} {\bibfnamefont
  {E.}~\bibnamefont {Kammerloher}}, \bibinfo {author} {\bibfnamefont
  {R.}~\bibnamefont {Otten}}, \bibinfo {author} {\bibfnamefont
  {I.}~\bibnamefont {Seidler}}, \bibinfo {author} {\bibfnamefont
  {R.}~\bibnamefont {Xue}}, \bibinfo {author} {\bibfnamefont {L.~R.}\
  \bibnamefont {Schreiber}},\ and\ \bibinfo {author} {\bibfnamefont
  {H.}~\bibnamefont {Bluhm}},\ }\href@noop {} {\bibinfo {title} {in
  preparation}} (\bibinfo {year} {2023})\BibitemShut {NoStop}%
\bibitem [{\citenamefont {McNeil}\ \emph {et~al.}(2011)\citenamefont {McNeil},
  \citenamefont {Kataoka}, \citenamefont {Ford}, \citenamefont {Barnes},
  \citenamefont {Anderson}, \citenamefont {Jones}, \citenamefont {Farrer},\
  and\ \citenamefont {Ritchie}}]{McNeil2011}%
  \BibitemOpen
  \bibfield  {author} {\bibinfo {author} {\bibfnamefont {R.~P.~G.}\
  \bibnamefont {McNeil}}, \bibinfo {author} {\bibfnamefont {M.}~\bibnamefont
  {Kataoka}}, \bibinfo {author} {\bibfnamefont {C.~J.~B.}\ \bibnamefont
  {Ford}}, \bibinfo {author} {\bibfnamefont {C.~H.~W.}\ \bibnamefont {Barnes}},
  \bibinfo {author} {\bibfnamefont {D.}~\bibnamefont {Anderson}}, \bibinfo
  {author} {\bibfnamefont {G.~A.~C.}\ \bibnamefont {Jones}}, \bibinfo {author}
  {\bibfnamefont {I.}~\bibnamefont {Farrer}},\ and\ \bibinfo {author}
  {\bibfnamefont {D.~A.}\ \bibnamefont {Ritchie}},\ }\bibfield  {title}
  {\bibinfo {title} {On-demand single-electron transfer between distant quantum
  dots},\ }\href {https://doi.org/10.1038/nature10444} {\bibfield  {journal}
  {\bibinfo  {journal} {Nature}\ }\textbf {\bibinfo {volume} {477}},\ \bibinfo
  {pages} {439} (\bibinfo {year} {2011})}\BibitemShut {NoStop}%
\bibitem [{\citenamefont {Takada}\ \emph {et~al.}(2019)\citenamefont {Takada},
  \citenamefont {Edlbauer}, \citenamefont {Lepage}, \citenamefont {Wang},
  \citenamefont {Mortemousque}, \citenamefont {Georgiou}, \citenamefont
  {Barnes}, \citenamefont {Ford}, \citenamefont {Yuan}, \citenamefont {Santos},
  \citenamefont {Waintal}, \citenamefont {Ludwig}, \citenamefont {Wieck},
  \citenamefont {Urdampilleta}, \citenamefont {Meunier},\ and\ \citenamefont
  {B{\"a}uerle}}]{Takada2019}%
  \BibitemOpen
  \bibfield  {author} {\bibinfo {author} {\bibfnamefont {S.}~\bibnamefont
  {Takada}}, \bibinfo {author} {\bibfnamefont {H.}~\bibnamefont {Edlbauer}},
  \bibinfo {author} {\bibfnamefont {H.~V.}\ \bibnamefont {Lepage}}, \bibinfo
  {author} {\bibfnamefont {J.}~\bibnamefont {Wang}}, \bibinfo {author}
  {\bibfnamefont {P.-A.}\ \bibnamefont {Mortemousque}}, \bibinfo {author}
  {\bibfnamefont {G.}~\bibnamefont {Georgiou}}, \bibinfo {author}
  {\bibfnamefont {C.~H.~W.}\ \bibnamefont {Barnes}}, \bibinfo {author}
  {\bibfnamefont {C.~J.~B.}\ \bibnamefont {Ford}}, \bibinfo {author}
  {\bibfnamefont {M.}~\bibnamefont {Yuan}}, \bibinfo {author} {\bibfnamefont
  {P.~V.}\ \bibnamefont {Santos}}, \bibinfo {author} {\bibfnamefont
  {X.}~\bibnamefont {Waintal}}, \bibinfo {author} {\bibfnamefont
  {A.}~\bibnamefont {Ludwig}}, \bibinfo {author} {\bibfnamefont {A.~D.}\
  \bibnamefont {Wieck}}, \bibinfo {author} {\bibfnamefont {M.}~\bibnamefont
  {Urdampilleta}}, \bibinfo {author} {\bibfnamefont {T.}~\bibnamefont
  {Meunier}},\ and\ \bibinfo {author} {\bibfnamefont {C.}~\bibnamefont
  {B{\"a}uerle}},\ }\bibfield  {title} {\bibinfo {title} {Sound-driven
  single-electron transfer in a circuit of coupled quantum rails},\ }\href
  {https://doi.org/10.1038/s41467-019-12514-w} {\bibfield  {journal} {\bibinfo
  {journal} {Nat Commun}\ }\textbf {\bibinfo {volume} {10}},\ \bibinfo {pages}
  {4557} (\bibinfo {year} {2019})}\BibitemShut {NoStop}%
\bibitem [{\citenamefont {Jadot}\ \emph {et~al.}(2021)\citenamefont {Jadot},
  \citenamefont {Mortemousque}, \citenamefont {Chanrion}, \citenamefont
  {Thiney}, \citenamefont {Ludwig}, \citenamefont {Wieck}, \citenamefont
  {Urdampilleta}, \citenamefont {Bäuerle},\ and\ \citenamefont
  {Meunier}}]{Jadot2021}%
  \BibitemOpen
  \bibfield  {author} {\bibinfo {author} {\bibfnamefont {B.}~\bibnamefont
  {Jadot}}, \bibinfo {author} {\bibfnamefont {P.-A.}\ \bibnamefont
  {Mortemousque}}, \bibinfo {author} {\bibfnamefont {E.}~\bibnamefont
  {Chanrion}}, \bibinfo {author} {\bibfnamefont {V.}~\bibnamefont {Thiney}},
  \bibinfo {author} {\bibfnamefont {A.}~\bibnamefont {Ludwig}}, \bibinfo
  {author} {\bibfnamefont {A.~D.}\ \bibnamefont {Wieck}}, \bibinfo {author}
  {\bibfnamefont {M.}~\bibnamefont {Urdampilleta}}, \bibinfo {author}
  {\bibfnamefont {C.}~\bibnamefont {Bäuerle}},\ and\ \bibinfo {author}
  {\bibfnamefont {T.}~\bibnamefont {Meunier}},\ }\bibfield  {title} {\bibinfo
  {title} {Distant spin entanglement via fast and coherent electron
  shuttling},\ }\href {https://doi.org/10.1038/s41565-021-00846-y} {\bibfield
  {journal} {\bibinfo  {journal} {Nature Nanotech}\ }\textbf {\bibinfo {volume}
  {16}},\ \bibinfo {pages} {570} (\bibinfo {year} {2021})}\BibitemShut
  {NoStop}%
\bibitem [{\citenamefont {Xue}\ \emph {et~al.}(2022)\citenamefont {Xue},
  \citenamefont {Russ}, \citenamefont {Samkharadze}, \citenamefont {Undseth},
  \citenamefont {Sammak}, \citenamefont {Scappucci},\ and\ \citenamefont
  {Vandersypen}}]{Xue2022}%
  \BibitemOpen
  \bibfield  {author} {\bibinfo {author} {\bibfnamefont {X.}~\bibnamefont
  {Xue}}, \bibinfo {author} {\bibfnamefont {M.}~\bibnamefont {Russ}}, \bibinfo
  {author} {\bibfnamefont {N.}~\bibnamefont {Samkharadze}}, \bibinfo {author}
  {\bibfnamefont {B.}~\bibnamefont {Undseth}}, \bibinfo {author} {\bibfnamefont
  {A.}~\bibnamefont {Sammak}}, \bibinfo {author} {\bibfnamefont
  {G.}~\bibnamefont {Scappucci}},\ and\ \bibinfo {author} {\bibfnamefont
  {L.~M.~K.}\ \bibnamefont {Vandersypen}},\ }\bibfield  {title} {\bibinfo
  {title} {Quantum logic with spin qubits crossing the surface code
  threshold},\ }\href {https://doi.org/10.1038/s41586-021-04273-w} {\bibfield
  {journal} {\bibinfo  {journal} {Nature}\ }\textbf {\bibinfo {volume} {601}},\
  \bibinfo {pages} {343} (\bibinfo {year} {2022})}\BibitemShut {NoStop}%
\bibitem [{\citenamefont {Noiri}\ \emph
  {et~al.}(2022{\natexlab{a}})\citenamefont {Noiri}, \citenamefont {Takeda},
  \citenamefont {Nakajima}, \citenamefont {Kobayashi}, \citenamefont {Sammak},
  \citenamefont {Scappucci},\ and\ \citenamefont {Tarucha}}]{Noiri2022_Nat}%
  \BibitemOpen
  \bibfield  {author} {\bibinfo {author} {\bibfnamefont {A.}~\bibnamefont
  {Noiri}}, \bibinfo {author} {\bibfnamefont {K.}~\bibnamefont {Takeda}},
  \bibinfo {author} {\bibfnamefont {T.}~\bibnamefont {Nakajima}}, \bibinfo
  {author} {\bibfnamefont {T.}~\bibnamefont {Kobayashi}}, \bibinfo {author}
  {\bibfnamefont {A.}~\bibnamefont {Sammak}}, \bibinfo {author} {\bibfnamefont
  {G.}~\bibnamefont {Scappucci}},\ and\ \bibinfo {author} {\bibfnamefont
  {S.}~\bibnamefont {Tarucha}},\ }\bibfield  {title} {\bibinfo {title} {Fast
  universal quantum gate above the fault-tolerance threshold in silicon},\
  }\href {https://doi.org/10.1038/s41586-021-04182-y} {\bibfield  {journal}
  {\bibinfo  {journal} {Nature}\ }\textbf {\bibinfo {volume} {601}},\ \bibinfo
  {pages} {338} (\bibinfo {year} {2022}{\natexlab{a}})}\BibitemShut {NoStop}%
\bibitem [{\citenamefont {Philips}\ \emph {et~al.}(2022)\citenamefont
  {Philips}, \citenamefont {M{\k{a}}dzik}, \citenamefont {Amitonov},
  \citenamefont {de~Snoo}, \citenamefont {Russ}, \citenamefont {Kalhor},
  \citenamefont {Volk}, \citenamefont {Lawrie}, \citenamefont {Brousse},
  \citenamefont {Tryputen}, \citenamefont {Wuetz}, \citenamefont {Sammak},
  \citenamefont {Veldhorst}, \citenamefont {Scappucci},\ and\ \citenamefont
  {Vandersypen}}]{Philips2022}%
  \BibitemOpen
  \bibfield  {author} {\bibinfo {author} {\bibfnamefont {S.~G.~J.}\
  \bibnamefont {Philips}}, \bibinfo {author} {\bibfnamefont {M.~T.}\
  \bibnamefont {M{\k{a}}dzik}}, \bibinfo {author} {\bibfnamefont {S.~V.}\
  \bibnamefont {Amitonov}}, \bibinfo {author} {\bibfnamefont {S.~L.}\
  \bibnamefont {de~Snoo}}, \bibinfo {author} {\bibfnamefont {M.}~\bibnamefont
  {Russ}}, \bibinfo {author} {\bibfnamefont {N.}~\bibnamefont {Kalhor}},
  \bibinfo {author} {\bibfnamefont {C.}~\bibnamefont {Volk}}, \bibinfo {author}
  {\bibfnamefont {W.~I.~L.}\ \bibnamefont {Lawrie}}, \bibinfo {author}
  {\bibfnamefont {D.}~\bibnamefont {Brousse}}, \bibinfo {author} {\bibfnamefont
  {L.}~\bibnamefont {Tryputen}}, \bibinfo {author} {\bibfnamefont {B.~P.}\
  \bibnamefont {Wuetz}}, \bibinfo {author} {\bibfnamefont {A.}~\bibnamefont
  {Sammak}}, \bibinfo {author} {\bibfnamefont {M.}~\bibnamefont {Veldhorst}},
  \bibinfo {author} {\bibfnamefont {G.}~\bibnamefont {Scappucci}},\ and\
  \bibinfo {author} {\bibfnamefont {L.~M.~K.}\ \bibnamefont {Vandersypen}},\
  }\bibfield  {title} {\bibinfo {title} {Universal control of a six-qubit
  quantum processor in silicon},\ }\href
  {https://doi.org/10.1038/s41586-022-05117-x} {\bibfield  {journal} {\bibinfo
  {journal} {Nature}\ }\textbf {\bibinfo {volume} {609}},\ \bibinfo {pages}
  {919} (\bibinfo {year} {2022})}\BibitemShut {NoStop}%
\bibitem [{\citenamefont {Mills}\ \emph {et~al.}(2022)\citenamefont {Mills},
  \citenamefont {Guinn}, \citenamefont {Gullans}, \citenamefont {Sigillito},
  \citenamefont {Feldman}, \citenamefont {Nielsen},\ and\ \citenamefont
  {Petta}}]{Mills2022}%
  \BibitemOpen
  \bibfield  {author} {\bibinfo {author} {\bibfnamefont {A.~R.}\ \bibnamefont
  {Mills}}, \bibinfo {author} {\bibfnamefont {C.~R.}\ \bibnamefont {Guinn}},
  \bibinfo {author} {\bibfnamefont {M.~J.}\ \bibnamefont {Gullans}}, \bibinfo
  {author} {\bibfnamefont {A.~J.}\ \bibnamefont {Sigillito}}, \bibinfo {author}
  {\bibfnamefont {M.~M.}\ \bibnamefont {Feldman}}, \bibinfo {author}
  {\bibfnamefont {E.}~\bibnamefont {Nielsen}},\ and\ \bibinfo {author}
  {\bibfnamefont {J.~R.}\ \bibnamefont {Petta}},\ }\bibfield  {title} {\bibinfo
  {title} {Two-qubit silicon quantum processor with operation fidelity
  exceeding 99\%},\ }\href {https://doi.org/10.1126/sciadv.abn5130} {\bibfield
  {journal} {\bibinfo  {journal} {Sci. Adv.}\ }\textbf {\bibinfo {volume}
  {8}},\ \bibinfo {pages} {eabn5130} (\bibinfo {year} {2022})}\BibitemShut
  {NoStop}%
\bibitem [{\citenamefont {Bertrand}\ \emph {et~al.}(2016)\citenamefont
  {Bertrand}, \citenamefont {Hermelin}, \citenamefont {Takada}, \citenamefont
  {Yamamoto}, \citenamefont {Tarucha}, \citenamefont {Ludwig}, \citenamefont
  {Wieck}, \citenamefont {Bäuerle},\ and\ \citenamefont
  {Meunier}}]{Bertrand2016}%
  \BibitemOpen
  \bibfield  {author} {\bibinfo {author} {\bibfnamefont {B.}~\bibnamefont
  {Bertrand}}, \bibinfo {author} {\bibfnamefont {S.}~\bibnamefont {Hermelin}},
  \bibinfo {author} {\bibfnamefont {S.}~\bibnamefont {Takada}}, \bibinfo
  {author} {\bibfnamefont {M.}~\bibnamefont {Yamamoto}}, \bibinfo {author}
  {\bibfnamefont {S.}~\bibnamefont {Tarucha}}, \bibinfo {author} {\bibfnamefont
  {A.}~\bibnamefont {Ludwig}}, \bibinfo {author} {\bibfnamefont {A.~D.}\
  \bibnamefont {Wieck}}, \bibinfo {author} {\bibfnamefont {C.}~\bibnamefont
  {Bäuerle}},\ and\ \bibinfo {author} {\bibfnamefont {T.}~\bibnamefont
  {Meunier}},\ }\bibfield  {title} {\bibinfo {title} {Fast spin information
  transfer between distant quantum dots using individual electrons},\ }\href
  {https://doi.org/10.1038/nnano.2016.82} {\bibfield  {journal} {\bibinfo
  {journal} {Nature Nanotech}\ }\textbf {\bibinfo {volume} {11}},\ \bibinfo
  {pages} {672} (\bibinfo {year} {2016})}\BibitemShut {NoStop}%
\bibitem [{\citenamefont {Mills}\ \emph {et~al.}(2019)\citenamefont {Mills},
  \citenamefont {Zajac}, \citenamefont {Gullans}, \citenamefont {Schupp},
  \citenamefont {Hazard},\ and\ \citenamefont {Petta}}]{Mills2019}%
  \BibitemOpen
  \bibfield  {author} {\bibinfo {author} {\bibfnamefont {A.~R.}\ \bibnamefont
  {Mills}}, \bibinfo {author} {\bibfnamefont {D.~M.}\ \bibnamefont {Zajac}},
  \bibinfo {author} {\bibfnamefont {M.~J.}\ \bibnamefont {Gullans}}, \bibinfo
  {author} {\bibfnamefont {F.~J.}\ \bibnamefont {Schupp}}, \bibinfo {author}
  {\bibfnamefont {T.~M.}\ \bibnamefont {Hazard}},\ and\ \bibinfo {author}
  {\bibfnamefont {J.~R.}\ \bibnamefont {Petta}},\ }\bibfield  {title} {\bibinfo
  {title} {Shuttling a single charge across a one-dimensional array of silicon
  quantum dots},\ }\href {https://doi.org/10.1038/s41467-019-08970-z}
  {\bibfield  {journal} {\bibinfo  {journal} {Nat Commun}\ }\textbf {\bibinfo
  {volume} {10}},\ \bibinfo {pages} {1063} (\bibinfo {year}
  {2019})}\BibitemShut {NoStop}%
\bibitem [{\citenamefont {Noiri}\ \emph
  {et~al.}(2022{\natexlab{b}})\citenamefont {Noiri}, \citenamefont {Takeda},
  \citenamefont {Nakajima}, \citenamefont {Kobayashi}, \citenamefont {Sammak},
  \citenamefont {Scappucci},\ and\ \citenamefont
  {Tarucha}}]{Noiri2022_NatComm}%
  \BibitemOpen
  \bibfield  {author} {\bibinfo {author} {\bibfnamefont {A.}~\bibnamefont
  {Noiri}}, \bibinfo {author} {\bibfnamefont {K.}~\bibnamefont {Takeda}},
  \bibinfo {author} {\bibfnamefont {T.}~\bibnamefont {Nakajima}}, \bibinfo
  {author} {\bibfnamefont {T.}~\bibnamefont {Kobayashi}}, \bibinfo {author}
  {\bibfnamefont {A.}~\bibnamefont {Sammak}}, \bibinfo {author} {\bibfnamefont
  {G.}~\bibnamefont {Scappucci}},\ and\ \bibinfo {author} {\bibfnamefont
  {S.}~\bibnamefont {Tarucha}},\ }\bibfield  {title} {\bibinfo {title} {A
  shuttling-based two-qubit logic gate for linking distant silicon quantum
  processors},\ }\href {https://doi.org/10.1038/s41467-022-33453-z} {\bibfield
  {journal} {\bibinfo  {journal} {Nat Commun}\ }\textbf {\bibinfo {volume}
  {13}},\ \bibinfo {pages} {5740} (\bibinfo {year}
  {2022}{\natexlab{b}})}\BibitemShut {NoStop}%
\bibitem [{\citenamefont {Zwerver}\ \emph {et~al.}(2022)\citenamefont
  {Zwerver}, \citenamefont {Amitonov}, \citenamefont {de~Snoo}, \citenamefont
  {Mądzik}, \citenamefont {Russ}, \citenamefont {Sammak}, \citenamefont
  {Scappucci},\ and\ \citenamefont {Vandersypen}}]{Zwerver2022}%
  \BibitemOpen
  \bibfield  {author} {\bibinfo {author} {\bibfnamefont {A.~M.~J.}\
  \bibnamefont {Zwerver}}, \bibinfo {author} {\bibfnamefont {S.~V.}\
  \bibnamefont {Amitonov}}, \bibinfo {author} {\bibfnamefont {S.~L.}\
  \bibnamefont {de~Snoo}}, \bibinfo {author} {\bibfnamefont {M.~T.}\
  \bibnamefont {Mądzik}}, \bibinfo {author} {\bibfnamefont {M.}~\bibnamefont
  {Russ}}, \bibinfo {author} {\bibfnamefont {A.}~\bibnamefont {Sammak}},
  \bibinfo {author} {\bibfnamefont {G.}~\bibnamefont {Scappucci}},\ and\
  \bibinfo {author} {\bibfnamefont {L.~M.~K.}\ \bibnamefont {Vandersypen}},\
  }\href@noop {} {\bibinfo {title} {Shuttling an electron spin through a
  silicon quantum dot array}} (\bibinfo {year} {2022}),\ \Eprint
  {https://arxiv.org/abs/2209.00920} {arXiv:2209.00920} \BibitemShut {NoStop}%
\bibitem [{\citenamefont {Feng}\ \emph {et~al.}(2023)\citenamefont {Feng},
  \citenamefont {Yoneda}, \citenamefont {Huang}, \citenamefont {Su},
  \citenamefont {Tanttu}, \citenamefont {Yang}, \citenamefont {Cifuentes},
  \citenamefont {Chan}, \citenamefont {Gilbert}, \citenamefont {Leon},
  \citenamefont {Hudson}, \citenamefont {Itoh}, \citenamefont {Laucht},
  \citenamefont {Dzurak},\ and\ \citenamefont {Saraiva}}]{Feng2023}%
  \BibitemOpen
  \bibfield  {author} {\bibinfo {author} {\bibfnamefont {M.}~\bibnamefont
  {Feng}}, \bibinfo {author} {\bibfnamefont {J.}~\bibnamefont {Yoneda}},
  \bibinfo {author} {\bibfnamefont {W.}~\bibnamefont {Huang}}, \bibinfo
  {author} {\bibfnamefont {Y.}~\bibnamefont {Su}}, \bibinfo {author}
  {\bibfnamefont {T.}~\bibnamefont {Tanttu}}, \bibinfo {author} {\bibfnamefont
  {C.~H.}\ \bibnamefont {Yang}}, \bibinfo {author} {\bibfnamefont {J.~D.}\
  \bibnamefont {Cifuentes}}, \bibinfo {author} {\bibfnamefont {K.~W.}\
  \bibnamefont {Chan}}, \bibinfo {author} {\bibfnamefont {W.}~\bibnamefont
  {Gilbert}}, \bibinfo {author} {\bibfnamefont {R.~C.~C.}\ \bibnamefont
  {Leon}}, \bibinfo {author} {\bibfnamefont {F.~E.}\ \bibnamefont {Hudson}},
  \bibinfo {author} {\bibfnamefont {K.~M.}\ \bibnamefont {Itoh}}, \bibinfo
  {author} {\bibfnamefont {A.}~\bibnamefont {Laucht}}, \bibinfo {author}
  {\bibfnamefont {A.~S.}\ \bibnamefont {Dzurak}},\ and\ \bibinfo {author}
  {\bibfnamefont {A.}~\bibnamefont {Saraiva}},\ }\bibfield  {title} {\bibinfo
  {title} {Control of dephasing in spin qubits during coherent transport in
  silicon},\ }\href {https://doi.org/10.1103/PhysRevB.107.085427} {\bibfield
  {journal} {\bibinfo  {journal} {Phys. Rev. B}\ }\textbf {\bibinfo {volume}
  {107}},\ \bibinfo {pages} {085427} (\bibinfo {year} {2023})}\BibitemShut
  {NoStop}%
\bibitem [{\citenamefont {Struck}\ \emph {et~al.}(2023)\citenamefont {Struck},
  \citenamefont {Volmer}, \citenamefont {Visser}, \citenamefont {Offermann},
  \citenamefont {Xue}, \citenamefont {Tu}, \citenamefont {Trellenkamp},
  \citenamefont {Bluhm},\ and\ \citenamefont {Schreiber}}]{Struck2023}%
  \BibitemOpen
  \bibfield  {author} {\bibinfo {author} {\bibfnamefont {T.}~\bibnamefont
  {Struck}}, \bibinfo {author} {\bibfnamefont {M.}~\bibnamefont {Volmer}},
  \bibinfo {author} {\bibfnamefont {L.}~\bibnamefont {Visser}}, \bibinfo
  {author} {\bibfnamefont {T.}~\bibnamefont {Offermann}}, \bibinfo {author}
  {\bibfnamefont {R.}~\bibnamefont {Xue}}, \bibinfo {author} {\bibfnamefont
  {J.-S.}\ \bibnamefont {Tu}}, \bibinfo {author} {\bibfnamefont
  {S.}~\bibnamefont {Trellenkamp}}, \bibinfo {author} {\bibfnamefont
  {H.}~\bibnamefont {Bluhm}},\ and\ \bibinfo {author} {\bibfnamefont {L.~R.}\
  \bibnamefont {Schreiber}},\ }\href@noop {} {\bibinfo {title} {in
  preparation}} (\bibinfo {year} {2023})\BibitemShut {NoStop}%
\bibitem [{\citenamefont {Klos}\ \emph {et~al.}(2018)\citenamefont {Klos},
  \citenamefont {Hassler}, \citenamefont {Cerfontaine}, \citenamefont {Bluhm},\
  and\ \citenamefont {Schreiber}}]{Klos2018}%
  \BibitemOpen
  \bibfield  {author} {\bibinfo {author} {\bibfnamefont {J.}~\bibnamefont
  {Klos}}, \bibinfo {author} {\bibfnamefont {F.}~\bibnamefont {Hassler}},
  \bibinfo {author} {\bibfnamefont {P.}~\bibnamefont {Cerfontaine}}, \bibinfo
  {author} {\bibfnamefont {H.}~\bibnamefont {Bluhm}},\ and\ \bibinfo {author}
  {\bibfnamefont {L.~R.}\ \bibnamefont {Schreiber}},\ }\bibfield  {title}
  {\bibinfo {title} {Calculation of tunnel couplings in open gate-defined
  disordered quantum dot systems},\ }\href
  {https://link.aps.org/doi/10.1103/PhysRevB.98.155320} {\bibfield  {journal}
  {\bibinfo  {journal} {Phys. Rev. B}\ }\textbf {\bibinfo {volume} {98}},\
  \bibinfo {pages} {155320} (\bibinfo {year} {2018})}\BibitemShut {NoStop}%
\bibitem [{\citenamefont {Humpohl}\ \emph {et~al.}(2023)\citenamefont
  {Humpohl}, \citenamefont {Prediger}, \citenamefont {Cerfontaine},
  \citenamefont {Bethke}, \citenamefont {Willmes}, \citenamefont {Meyer},
  \citenamefont {Hangleiter}, \citenamefont {Kammerloher}, \citenamefont
  {Lankes}, \citenamefont {Struck}, \citenamefont {Eendebak},\ and\
  \citenamefont {Xue}}]{Humpohl2023}%
  \BibitemOpen
  \bibfield  {author} {\bibinfo {author} {\bibfnamefont {S.}~\bibnamefont
  {Humpohl}}, \bibinfo {author} {\bibfnamefont {L.}~\bibnamefont {Prediger}},
  \bibinfo {author} {\bibfnamefont {P.}~\bibnamefont {Cerfontaine}}, \bibinfo
  {author} {\bibfnamefont {P.}~\bibnamefont {Bethke}}, \bibinfo {author}
  {\bibfnamefont {A.}~\bibnamefont {Willmes}}, \bibinfo {author} {\bibfnamefont
  {M.}~\bibnamefont {Meyer}}, \bibinfo {author} {\bibfnamefont
  {T.}~\bibnamefont {Hangleiter}}, \bibinfo {author} {\bibfnamefont
  {E.}~\bibnamefont {Kammerloher}}, \bibinfo {author} {\bibfnamefont
  {L.}~\bibnamefont {Lankes}}, \bibinfo {author} {\bibfnamefont
  {T.}~\bibnamefont {Struck}}, \bibinfo {author} {\bibfnamefont
  {P.}~\bibnamefont {Eendebak}},\ and\ \bibinfo {author} {\bibfnamefont
  {R.}~\bibnamefont {Xue}},\ }\href {https://doi.org/10.5281/zenodo.7777634}
  {\bibinfo {title} {qutech/qupulse: Release 0.8}} (\bibinfo {year}
  {2023})\BibitemShut {NoStop}%
\bibitem [{\citenamefont {Albrecht}\ \emph {et~al.}(2017)\citenamefont
  {Albrecht}, \citenamefont {Moers},\ and\ \citenamefont
  {Hermanns}}]{albrecht_hnf_2017}%
  \BibitemOpen
  \bibfield  {author} {\bibinfo {author} {\bibfnamefont {W.}~\bibnamefont
  {Albrecht}}, \bibinfo {author} {\bibfnamefont {J.}~\bibnamefont {Moers}},\
  and\ \bibinfo {author} {\bibfnamefont {B.}~\bibnamefont {Hermanns}},\
  }\bibfield  {title} {\bibinfo {title} {{HNF} - {Helmholtz} {Nano}
  {Facility}},\ }\href {https://doi.org/10.17815/jlsrf-3-158} {\bibfield
  {journal} {\bibinfo  {journal} {Journal of large-scale research facilities}\
  }\textbf {\bibinfo {volume} {3}},\ \bibinfo {pages} {A112} (\bibinfo {year}
  {2017})}\BibitemShut {NoStop}%
\end{thebibliography}
\end{document}